\documentclass[preprint,journal]{vgtc}       

\ifpdf
  \pdfoutput=1\relax                   
  \usepackage{graphicx}                
  \DeclareGraphicsExtensions{.pdf,.png,.jpg,.jpeg} 
\else
  \usepackage{graphicx}                
  \DeclareGraphicsExtensions{.eps}     
\fi%

\graphicspath{{figures/}{pictures/}{images/}{./}} 

\usepackage{microtype}                 
\PassOptionsToPackage{warn}{textcomp}  
\usepackage{textcomp}                  
\usepackage{mathptmx}                  
\usepackage{times}                     
\usepackage{cite}                      
\usepackage{tabu}                      
\usepackage{booktabs}                  

\usepackage{amsmath}
\usepackage{subfigure}
\usepackage{color}

\onlineid{1257}

\vgtccategory{Research}
\vgtcpapertype{algorithm/technique}

\title{
Animated Stickies: Fast Video Projection Mapping
onto a Markerless Plane through a
Direct Closed-Loop Alignment}

\author{Shingo Kagami, \textit{Member, IEEE}, and Koichi Hashimoto, \textit{Member, IEEE}}
\authorfooter{
\item
 Shingo Kagami is with the Graduate School of Information Sciences, Tohoku
 University. E-mail: swk(at)ic.is.tohoku.ac.jp.
\item
 Koichi Hashimoto is with the Graduate School of Information Sciences, Tohoku
 University. 
}

\shortauthortitle{Kagami \MakeLowercase{\textit{et al.}}: Animated Stickies: Fast Video Projection Mapping onto a Markerless Plane through a Direct Closed-Loop Alignment}

\abstract{This paper presents a fast projection mapping method for moving image
content projected onto a markerless planar surface using a low-latency Digital Micromirror Device (DMD)
projector. By adopting a closed-loop
alignment approach, in which not only the surface texture but also the
projected image is tracked by a camera, the proposed method is free
from a calibration or position adjustment between the camera and 
projector. We designed fiducial patterns to be inserted into a
fast flapping sequence of binary frames of the DMD projector, which
allows the simultaneous tracking of the surface texture and a fiducial
geometry separate from a single image captured by the camera. The
proposed method implemented on a CPU runs at 400 fps and enables
arbitrary video contents to be ``stuck'' onto a variety of textured
surfaces.%
} 


\CCScatlist{ 
 \CCScat{K.6.1}{Management of Computing and Information Systems}%
{Project and People Management}{Life Cycle};
 \CCScat{K.7.m}{The Computing Profession}{Miscellaneous}{Ethics}
}

\teaser{
\centering
\begin{minipage}{4.05cm} \centering
\includegraphics[width=4.0cm]{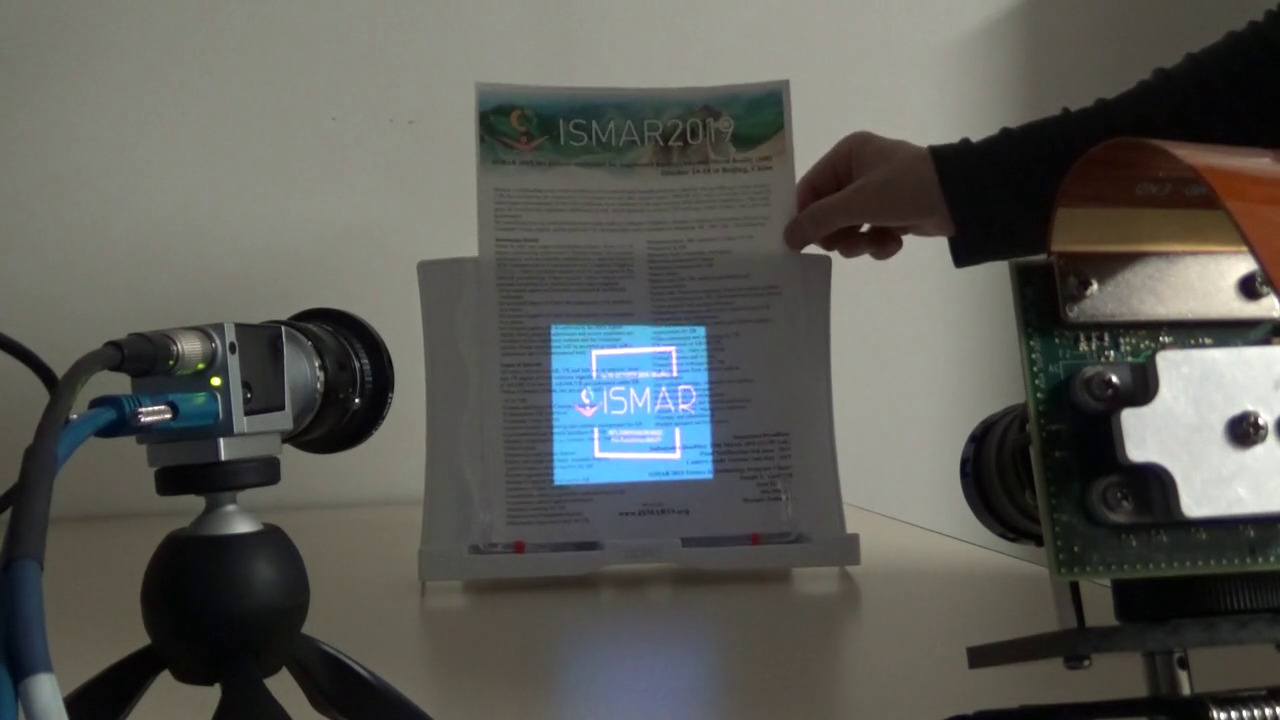}\\
{\footnotesize $t = 0.0 $ [s] }
\end{minipage}~%
\begin{minipage}{4.05cm} \centering
\includegraphics[width=4.0cm]{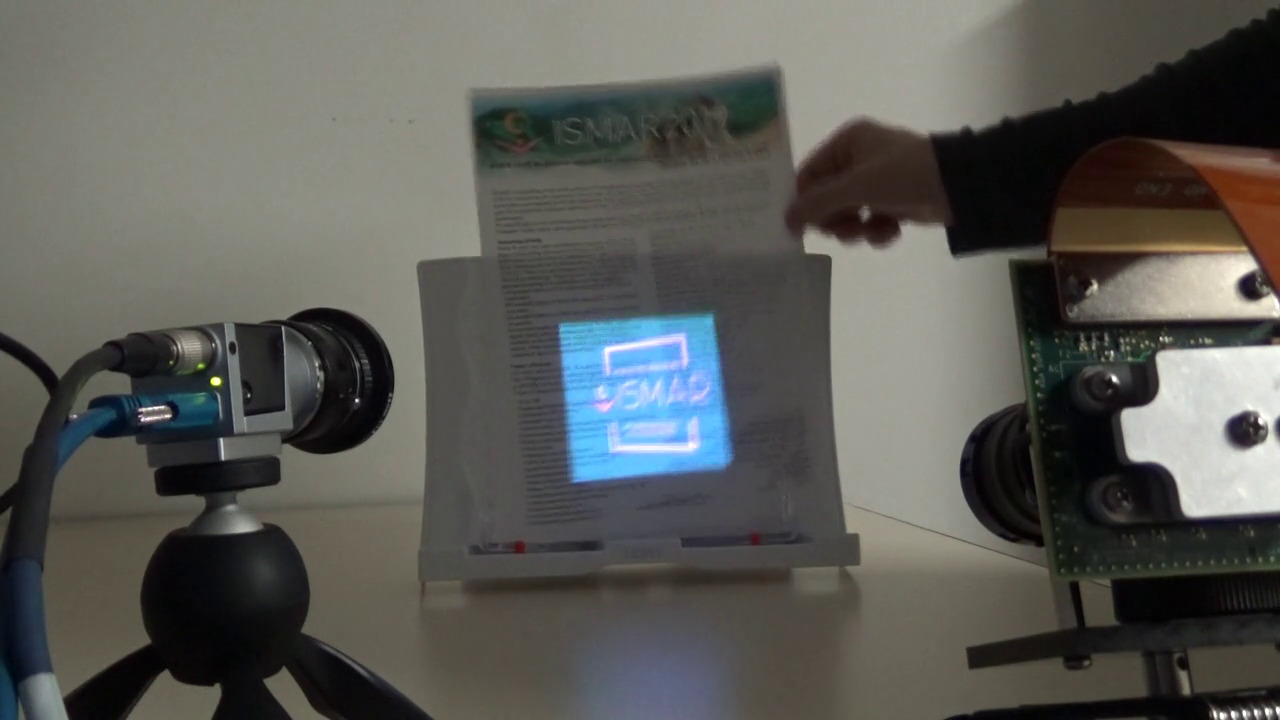}\\
{\footnotesize $t = 0.1 $ [s] }
\end{minipage}~%
\begin{minipage}{4.05cm} \centering
\includegraphics[width=4.0cm]{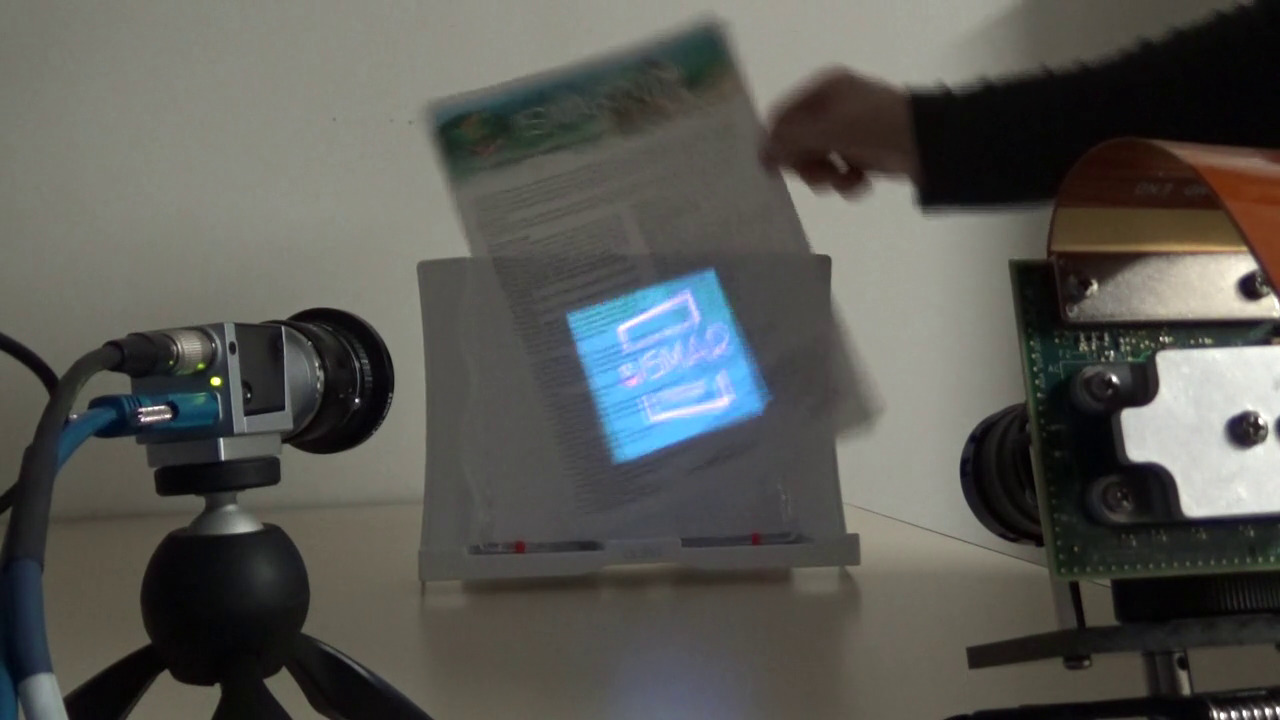}\\
{\footnotesize $t = 0.2 $ [s] }
\end{minipage}~%
\begin{minipage}{4.05cm} \centering
\includegraphics[width=4.0cm]{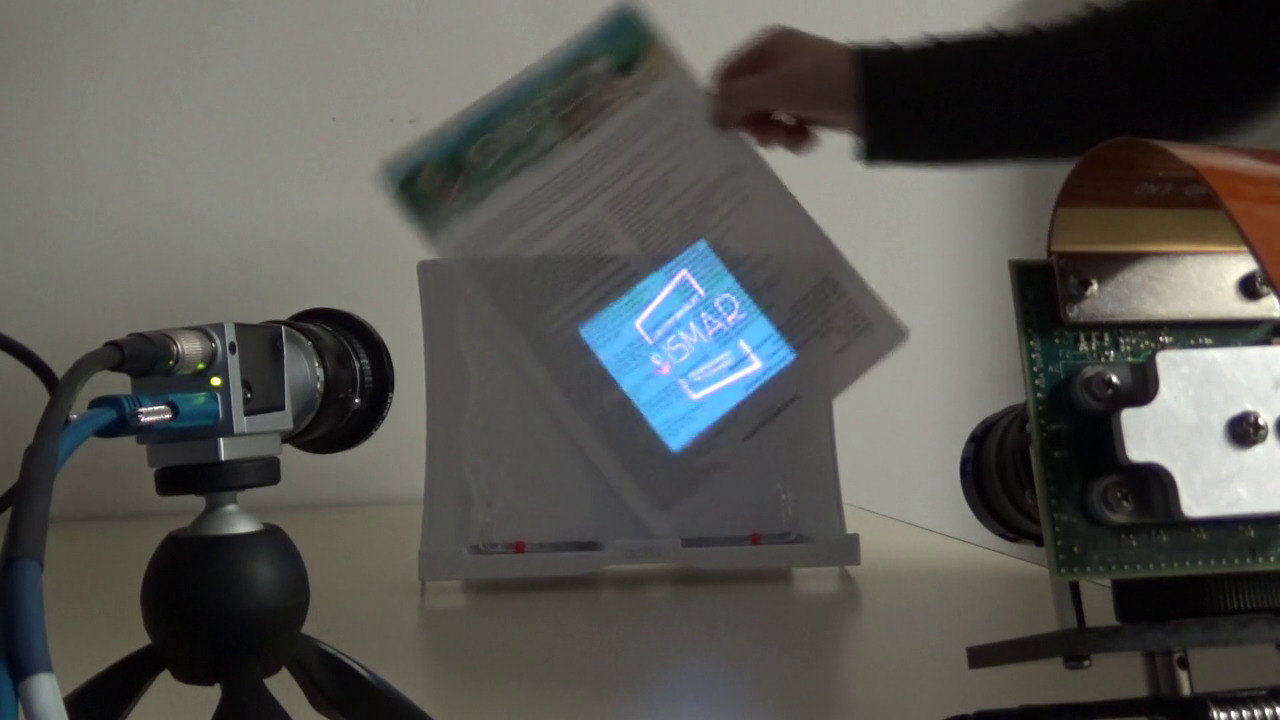}\\
{\footnotesize $t = 0.3 $ [s] }
\end{minipage}\\[2mm]
\begin{minipage}{4.05cm} \centering
\includegraphics[width=4.0cm]{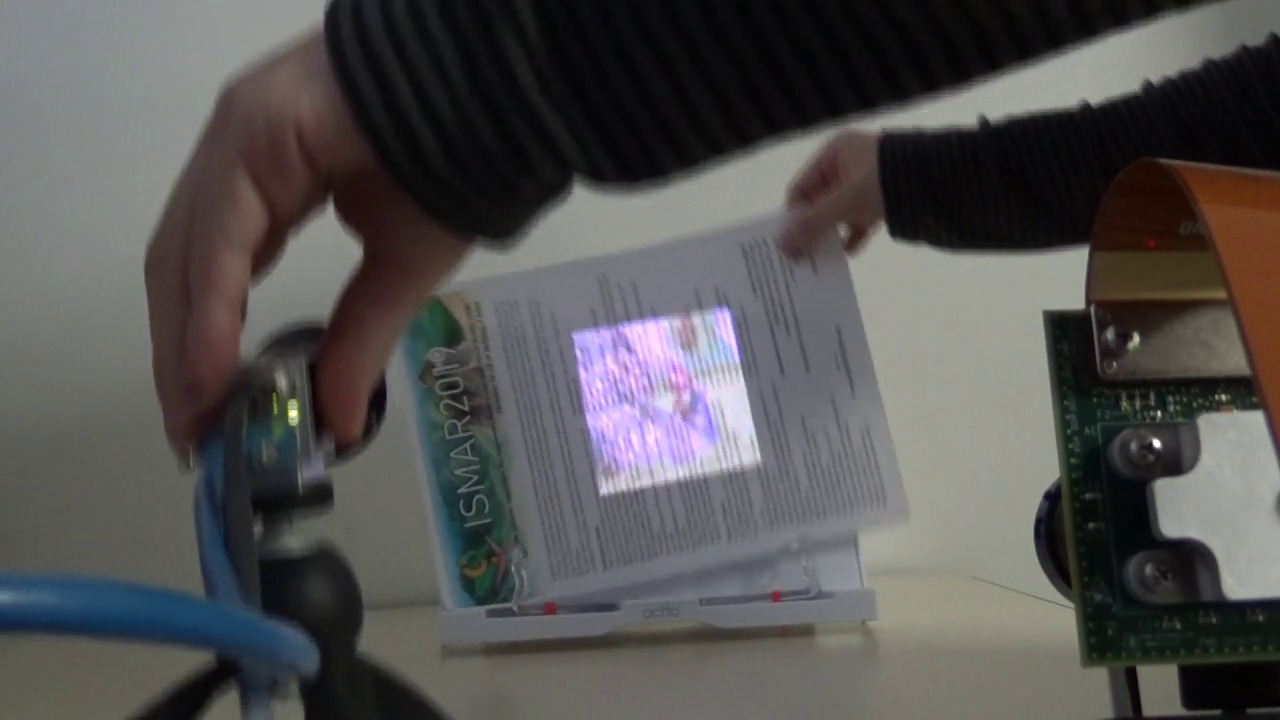}\\
{\footnotesize $t = 0.0 $ [s] }
\end{minipage}~%
\begin{minipage}{4.05cm} \centering
\includegraphics[width=4.0cm]{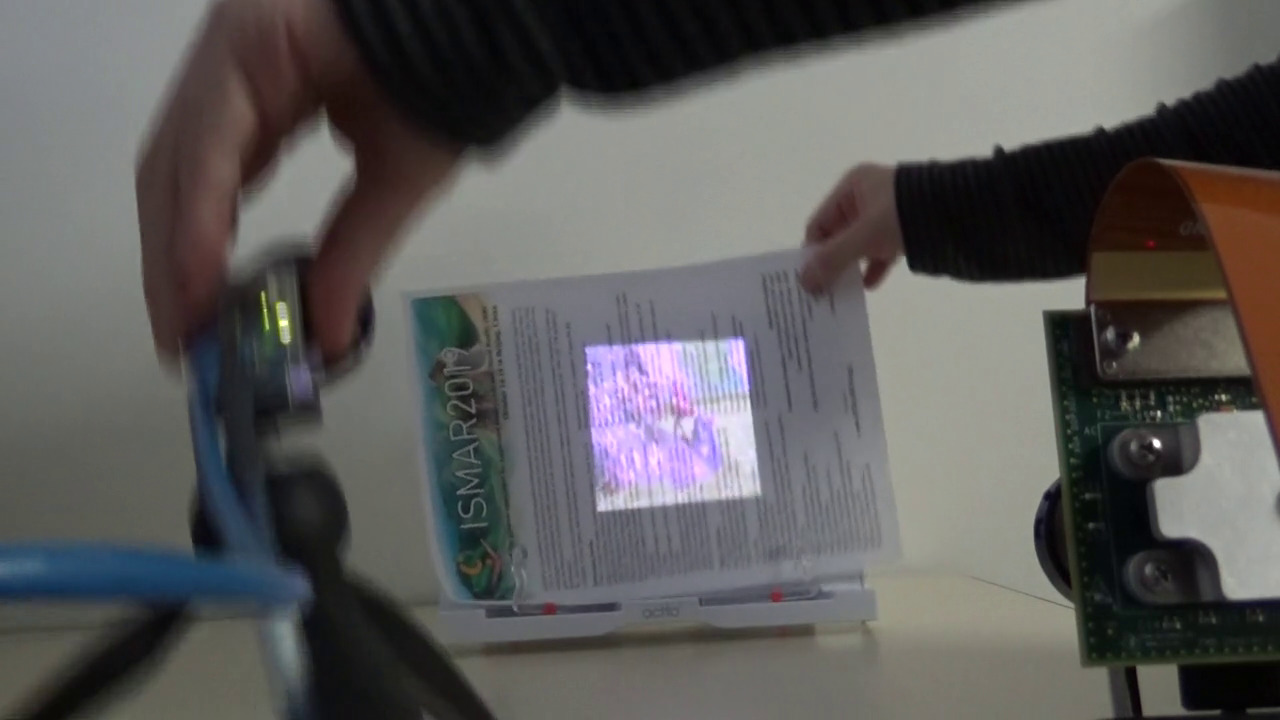}\\
{\footnotesize $t = 0.1 $ [s] }
\end{minipage}~%
\begin{minipage}{4.05cm} \centering
\includegraphics[width=4.0cm]{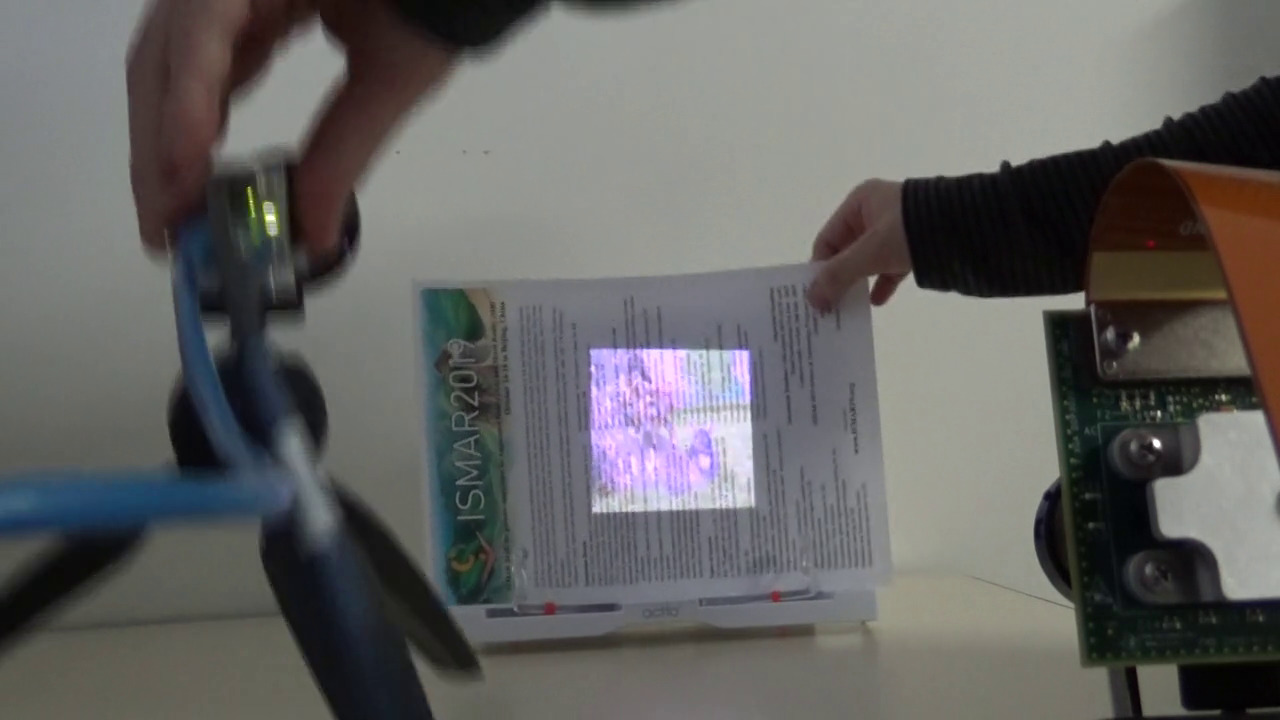}\\
{\footnotesize $t = 0.2 $ [s] }
\end{minipage}~%
\begin{minipage}{4.05cm} \centering
\includegraphics[width=4.0cm]{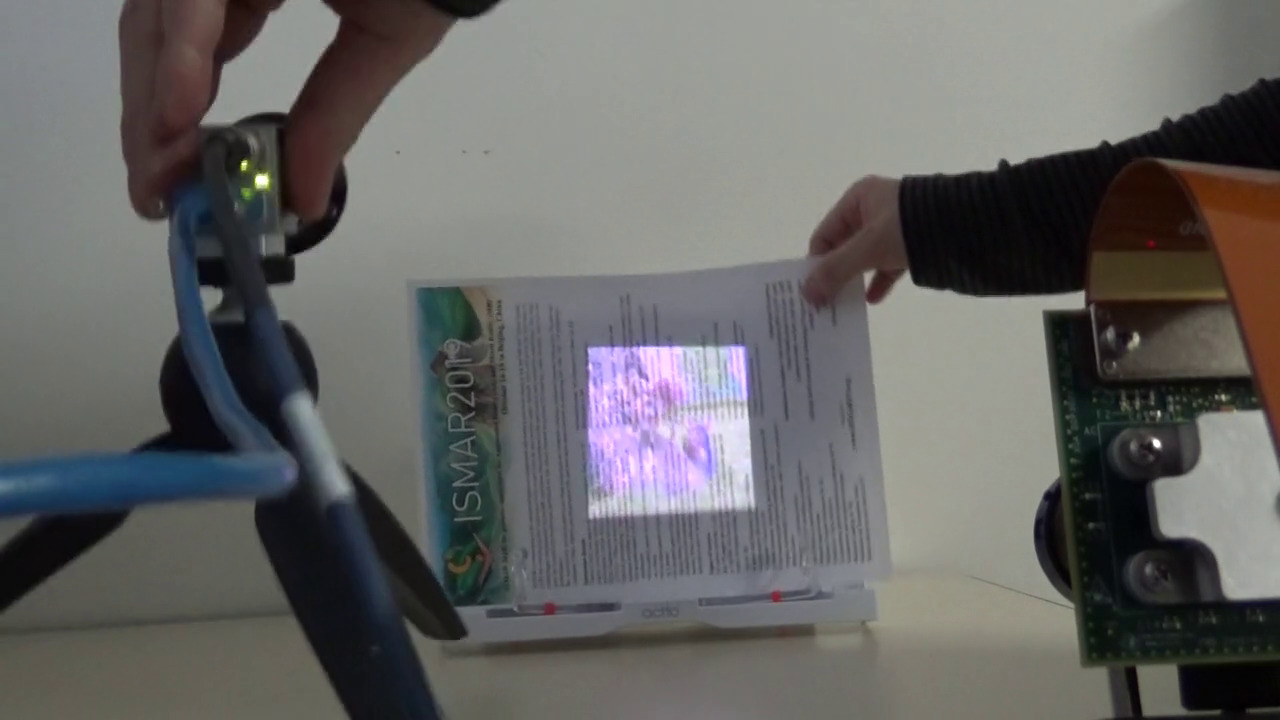}\\
{\footnotesize $t = 0.3 $ [s] }
\end{minipage}
\caption{Snapshot sequences of dynamic projection mapping onto fast
moving markerless surfaces. The bottom row indicates that moving the 
camera pose (left side in the photo) relative to the projector (right
side) does not disturb the mapping.}
\label{fig:cfp}
}

\vgtcinsertpkg

\begin{document}

\firstsection{Introduction}

\maketitle

The successful implementation of spatial augmented reality (SAR) in
dynamic scenarios necessitates the use of fast low-latency projection systems
that can adapt images quickly to scene motions.
State-of-the-art systems utilize Digital Micromirror Device (DMD)
projectors to update projected images at several hundreds of fps or
higher \cite{kagami2015, narita2017, bermano2017, kagami2018,
  miyashita2018}.

These systems may seem too costly today in terms of price, system size,
and operational cost to be used for consumer-level applications.
However, considering that DMD projectors in general are the most
popular choice for mobile and pico-projectors and that the core optical
engines of low-latency systems are the same as those used in
regular DMD products, the widespread use of commercial low-latency projectors in the future may be a possibility. Recalling
that some popular smartphones are already
equipped with high-speed (e.g., 240 fps) cameras, addressing the issue
of how to realize a casual consumer-level use of low-latency SAR systems
is important.

For such casual use, SAR systems should consist of a minimal number of
components and should be easy to set up. For example, it is desirable to avoid attaching markers on the target surfaces, as well as a precise adjustment of the camera-projector positions and
their calibration.

Image alignment techniques for augmented reality are generally
classified into two approaches: closed- and open-loop. Closed-loop
approaches \cite{johnson2007, audet2010, nakamura2012,
  zheng2013, naik2015, kagami2015, resch2016} track not only the
target surface but also the projected pattern to minimize
alignment errors, and are therefore insusceptible to, and occasionally
almost free from, positioning or calibration errors.

Most existing low-latency SAR systems employ open-loop approaches
to avoid projection pattern tracking, presumably owing to a
limited computation time budget. A popular tracking
measure is the use of infrared cameras, by which the projected patterns
are not observed. Narita et al.~\cite{narita2017} used markers drawn
on a surface using infrared-absorbing ink for tracking. Bermano et
al.~\cite{bermano2017} tracked a markerless human face illuminated by
infrared light. Both used coaxially aligned projector-camera
pairs allowing 3D position measurements to be avoided. However, the
alignment of the optical axes requires a careful operation and may not
be fit for casual use.

When we do not limit the discussion to cases using low-latency
projectors, the most popular choice for tracking in dynamic SAR
applications in the recent literature would be the use of depth sensors
\cite{siegl2015, zhou2016}. Because depth information is not affected by the projected
content, this approach is classified as open-loop unless other
sensory information is adjunctively used. Therefore, 
errors in 3D modeling and sensor calibration, for example, inevitably affect the
results. It is also challenging to achieve low-latency sensory
feedback with depth sensors when compared to the use of 2D
cameras. 

The closed-loop approach, by contrast, is a challenge owing to
the difficulty of target surface tracking under interference by the projected
content. A front door approach tackles this issue by incorporating the
effect of projection into the optimization process for
alignment purposes~\cite{audet2010, nakamura2012, audet2013, zheng2013}.
Adopting this approach to real-time tracking at a high frame rate,
however, is difficult mainly because the optimization problem becomes
complicated and intractable when the frame time is short.  
When a video
content, instead of a still image, is projected, the complexity increases
 because the template image for tracking changes
and must be initialized every frame. The dependency of the tracking accuracy
on the projection content is also problematic. When the tracker comes
across a featureless video frame, the tracking result will become
suddenly unstable.

Researchers occasionally have avoided this interference by limiting their target
to a texture-less solid-color surface. Johnson and
Fuchs~\cite{johnson2007} and Resch et al.~\cite{resch2016}
assumed a texture-less non-planar surface with a known shape, and
tracked the known feature points in the projected image reflected by the
surface to determine the projector pose. Kagami and
Hashimoto~\cite{kagami2015} assumed a white planar quadrangle surface, 
and tracked the projected image through a direct alignment and tracked
the four sides of the quadrangle by line fitting to the edges. In all
cases, the problem of dependency on the projection contents 
remains to be solved.

This problem of content dependency can be eliminated if we hide
imperceptible fiducial patterns in a projected video sequence, allowing the tracking of patterns to be geometrically equivalent to
the tracking of the projection content. A popular and classical approach to this is
to insert additional frames into the projection content \cite{raskar1998,
  cotting2004, mcdowall2004, hiraki2016, kusanagi2017}. Because DMD projectors
represent an image through a sequence of fast switching binary frames, it
is possible to make the inserted frames so short that the human eyes
can barely perceive them. 

However, even with this fiducial-hiding approach, we also need to face
the interference problem. For a static scene, it may be possible to
suppress the interference by processing multiple consecutive frames.
For example, by consecutively capturing a pattern frame and its
complementary-color frame, the pattern geometry can be extracted from
their difference \cite{grundhoefer2007, yamamoto2017} and possibly
recover the surface texture from their average. For a highly dynamic
scene, however, this will not work. The use of an infrared pattern
projection \cite{willis2011} or image steganographic method
\cite{suzuki2008} will not help in this regard.

In this light, we address the issue of tracking the surface texture and
projected fiducial patterns separately from a single camera frame.
Our approach is to make use of the spatial domain for separation under
the strong assumption of a high frame rate measurement. Thanks to
its closed-loop configuration, the proposed method is free from
a camera-projector calibration. 

This paper focuses on the planar target surfaces primarily
because the hardware platform we use \cite{kagami2018} supports only
a homography transformation with hard-wired logic implemented in the
projector. It should be noted, however, that focusing on the planar targets is
not only a simplification but incurs a peculiar difficulty and profound
significance. For example, the lack of 3D shape features prevents the use
of existing techniques relying on 3D information. From a practical
perspective, planar or approximately planar surfaces are in great
demand as projection targets.  
The possibility of generalization to non-planar surfaces
is discussed in \autoref{sec:limitations}.

\section{Projection System}

\begin{figure}[tb]
\centering
\includegraphics[width=\columnwidth]{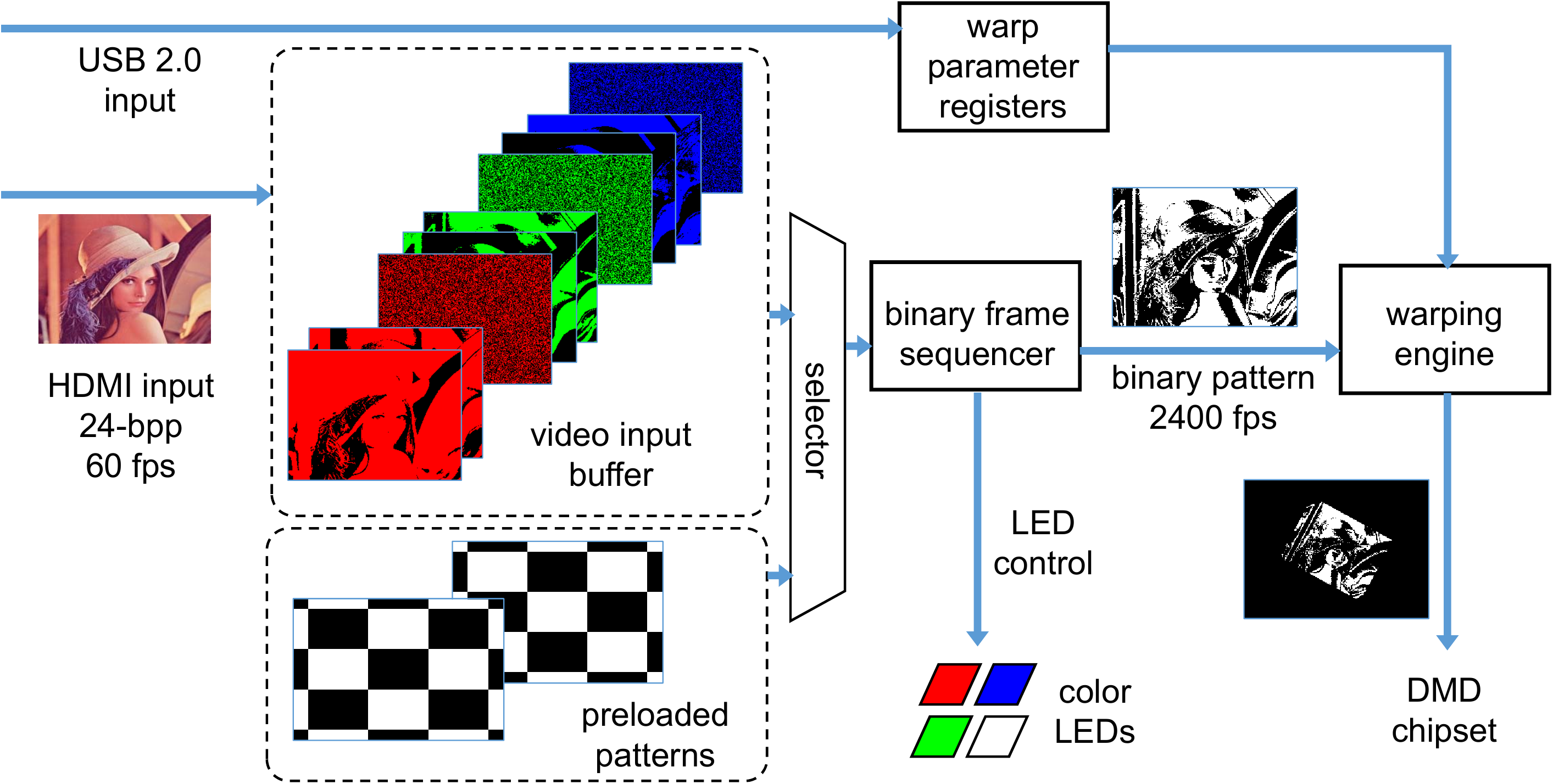}
\caption{Operation pipeline of our projection system.}
\label{fig:hardware_overview}
\end{figure}

This section describes the hardware platform \cite{kagami2018} upon
which the proposed method is implemented. This DMD projection system
has been designed to enable low-latency mapping of video-rate content
onto a moving surface without the need for generating content with a high frame rate. Note, however, that the proposed method itself will be
applicable to a wider class of high-speed projectors.

A DMD is a reflective spatial light modulator that produces monochrome
binary images at up to tens of kilohertz. Fast switching $2^n - 1$ binary
images are time-averaged by the human eye and perceived as an $n$-bit
gray-level image. With a light source whose brightness can be
modulated for each binary frame, the necessary number of binary frames
can be reduced, which is the principle that most high-speed DMD
projectors are based upon.

Our system adopts a different approach. Instead of increasing the
frame rate of the video content and thereby achieving a low-latency motion
adaptation, it warps each of the binary frames at the
binary frame rate. Oshiro et al.~\cite{oshiro2019} reported that
applying this technique to 60-fps images offers a perceptual image quality
comparable to that of high frame rate images. Similar approaches have
been proposed for head-mounted displays \cite{zheng2014, lincoln2016},
where binary frames are generated at every time instant; however, our approach is
simpler in that binary frames are simply selected instead of being
generated. Microsoft Hololens, using liquid crystal on silicon (LCoS)
instead of DMD, also applies a similar technique to each color field
\cite{klein2017}.

\autoref{fig:hardware_overview} shows the pipeline of the projection
system. It receives a 24-bpp 60-fps video stream through HDMI and
decomposes a frame into bitplanes.  Binary patterns are read out
at a binary frame rate of up to 2,470 fps from this input bitplane buffer
or a storage of preloaded binary patterns according to a predefined
sequence. A binary pattern is warped according to the homography
parameters that have been received most recently through a USB 2.0 port, and
sent to a 0.7" XGA DMD (Texas Instruments DLP7000BFLP). The sequencer
also controls RGB-White LEDs (Luminus SBM-40) that illuminate the DMD, and
the light reflected by the DMD travels through the projection optics (ViaLUX
STAR CORE-07).

It is notable that there is no need for high frame rate video input,
which leads to a low-cost implementation suitable for
consumer-level products. Another advantage of this approach is that
the binary pattern rate used to represent color images can be kept
relatively low (i.e., a few instead of tens of kilohertz),
which allows the use of a lower-cost DMD and simplified electronics design. 

\section{Fiducial Design and Plane Tracking Method}
\label{sec:fiducial_track}

\begin{figure}[tb]
\centering
\includegraphics[width=\columnwidth]{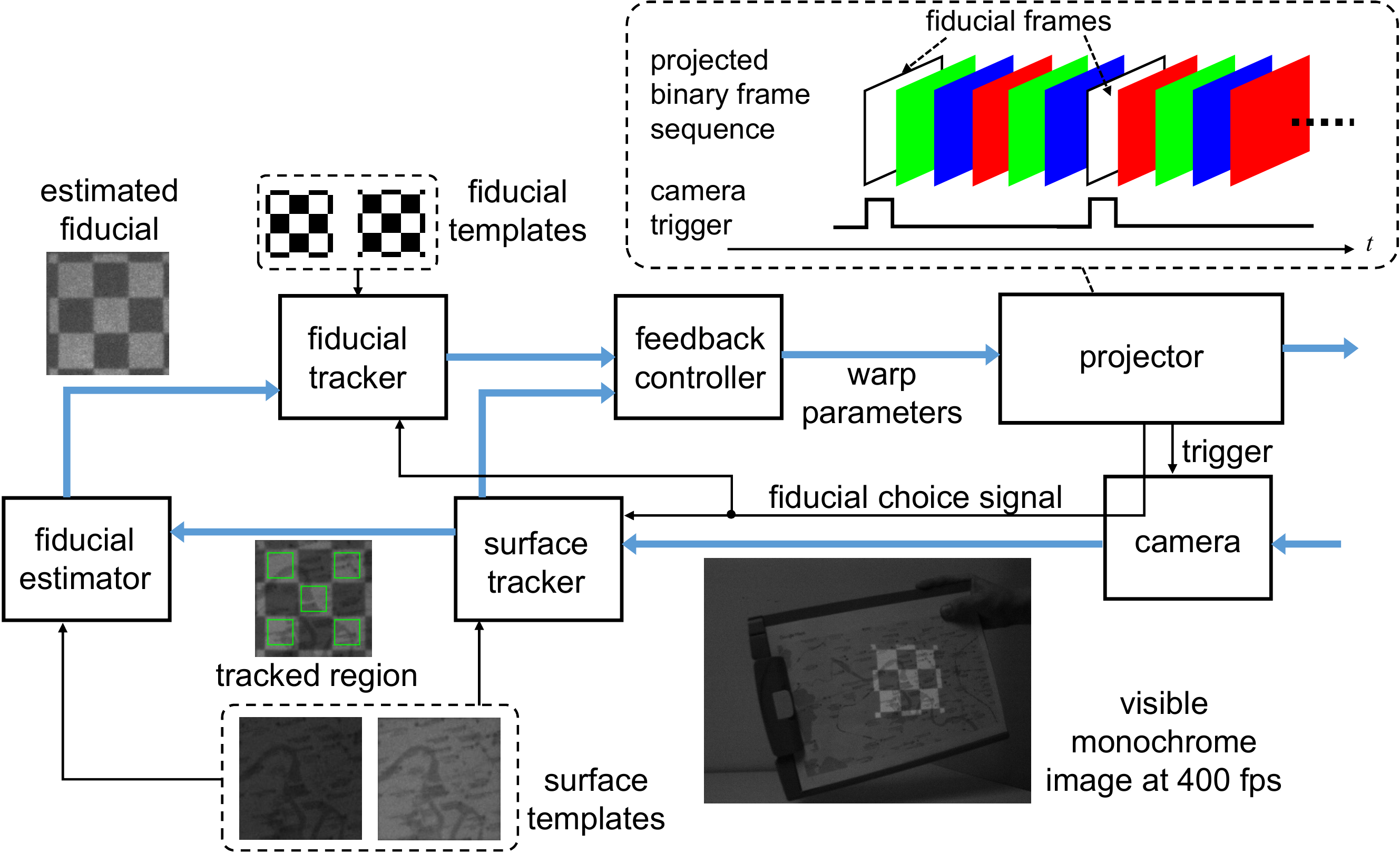}
\caption{Pipeline of the proposed method.}
\label{fig:algorithm_pipeline}
\end{figure}

\begin{figure}[tb]
\centering
\includegraphics[width=\columnwidth]{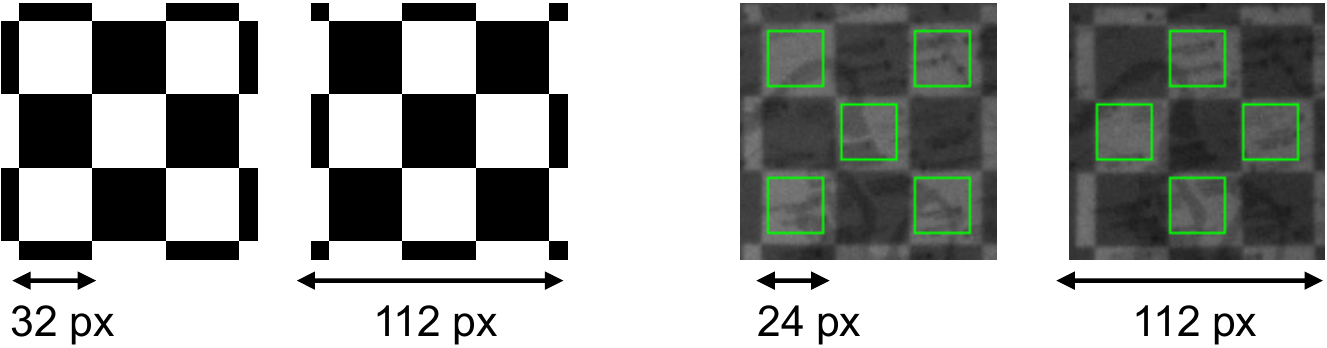}
\caption{Fiducial design (left) and corresponding area of interest of
  a surface (right) warped to the template size of the surface
  tracker. Only the pixels in the green rectangles are used in the direct
  alignment algorithm.}
\label{fig:fiducial_design}
\end{figure}

Fiducial patterns should be inserted into the binary frame sequence 
as frequently as required for sensor feedback, and should have the
following properties:
\begin{itemize}
\item imperceptibility (or hardly perceptible) by human eyes;
\item no disturbance of the surface texture tracking;
\item recoverability of geometry from a single frame measurement.
\end{itemize}

For the imperceptibility, the display period of the pattern must be
sufficiently short. When DMD projectors are used, it means the
pattern should be a binary or superposition of a few binary images. To avoid the time-integration of the patterns, which repeatedly appear at
the sensor feedback rate, from being visible, complementary-color pairs
of patterns should be displayed with equal occurrence frequency.

For surface tracking to be executed at a high frame rate with a
relatively low computational cost, we focus on direct alignment
methods, which directly minimize the differences of the pixel values without
an explicit feature extraction. It is well known that direct alignment
methods work well even if only a selected subset of the pixels are
used during optimization, the popular choices of which are a random selection or
high curvature texture points \cite{dellaert1999, baker2003}.

This motivates us to select camera image pixels illuminated by bright
pixels in the fiducial binary pattern displayed by the projector and
let them participate in the optimization. The issue faced here is that
it is unknown which pixels in the camera image are illuminated and
which pixels are not, which is the very goal of 
optimization.  

Therefore, we make a strong assumption that tracking in the previous
camera frame is successful and that inter-frame motion between the
consecutive camera images is sufficiently small thanks to high frame
rate measurements. We propose the use of chessboard-like patterns and
select those pixels that lie in {\it eroded} regions of white chessboard
cells in the previous camera frame. Because we assume that inter-frame
motion is small, we are sure that most of these pixels are also illuminated
by the white chessboard cells in the current camera frame. 

The next problem is to locate the fiducial pattern in the current
camera frame. Under the assumption that we have already tracked the
surface texture successfully, it is reasonable to consider the
reflectance of the surface at each camera pixel as equal
to that of the corresponding pixel 
measured when initialization of the tracking is conducted. This
information can be used to classify whether a pixel of interest is
illuminated by the chessboard white cell. 

More specifically, during the initialization phase, we store an image
$I^{+}$ of the tracked regions of the surface with all projector
pixels on, and another image $I^{-}$ with all pixels off,
and memorize the mean pixel values $\bar{I}^{+}$ and $\bar{I}^{-}$ of
both images.

During the tracking phase, we seek a uniform gain coefficient
multiplied with the pixel values in the tracked regions such that their
mean becomes equal to the memorized $\bar{I}^{+}$. After we obtain
the pixel-wise correspondence between the input image and the
memorized $I^{-}$ by completing the surface tracking, the pixel values
are normalized by dividing them by the corresponding pixel values in
$I^{-}$. This normalized image is now expected to have a value
close to $\bar{I}^{+} / \bar{I}^{-}$ at a pixel illuminated by the
chessboard white cells, and a considerably lower value otherwise. 
This normalized image is treated as an input to another direct
alignment process against the fiducial template image. Although this
normalization is only an approximation, it works sufficiently well for a direct
alignment against the binary-valued fiducial templates. 

The exception is a case in which the ambient light is too low, where
division by $I^{-}$ values is unreliable. In this case, namely, when 
$\bar{I}^{-}$ is less than 16 at an 8-bit level, 
our implementation simply uses the
gain-adapted surface tracking result as the input to the fiducial
tracking process. 

\autoref{fig:algorithm_pipeline} illustrates the pipeline of the
proposed method. We set the binary frame rate to 2,400~fps and insert
the fiducial patterns such that they appear in every sixth binary frame,
resulting in a 400-fps visual feedback rate. A trigger signal is sent
from the projector to a Basler USB-3 monochrome camera acA640-750um to
control the shutter, along with a binary signal indicating whether the
original fiducial pattern or its complementary pattern is presented.
Whereas the proposed method does not depend on a specific choice of
direct alignment methods, we chose the efficient second-order
minimization (ESM) method \cite{benhimane2007} for both surface and fiducial tracking
because it has been found to work well for the tracking of content
projected onto a plane
\cite{kagami2015}.

The design of the fiducial pattern is crucial in achieving high
accuracy tracking. In our initial attempt, we attempted the use of simple $3
\times 3$ chessboard patterns, although the resulting accuracy of the fiducial
positioning was far from satisfactory. Because only the borders of
the chessboard cells provide information to position the fiducial in the
image, they should appear as close as possible to the peripheries of
the video content area. By contrast, overly fine chessboard grids
make the surface tracking difficult. We chose a pattern that is
marginally larger than the $3 \times 3$ chessboard shown in
\autoref{fig:fiducial_design}.

The light source setting for the fiducial pattern projection is also
important. On the one hand, in order to keep the visible color
contrast of the presented content sufficiently high, the illumination
intensity for the fiducial should be as low as possible and the illumination period
should be as short as possible. On the other hand, the illumination
for the fiducial must provide a sufficient light amount for the
short-exposure camera measurement of the patterns to be reliable.
This tradeoff must be dealt with by considering many factors including
the surface reflectance and ambient light, and is manually adjusted in
our current implementation. A white light source color is chosen (or if a white light source is unavailable, all RGB values are turned on) because it is a neutral color for various types of spectral surface reflectance.

\section{Tracking Projection Algorithm}

\subsection{Initialization}

For initialization, a user-specified rectangle is given in the
projector image space. During the initialization phase, the projector
inserts a white frame, a black frame, an ArUco marker frame~\cite{garrido-jurado2014},
and an intensity-inverted counterpart of the ArUco
marker frame in place of the chessboard-like fiducials used in the
tracking phase, into the video sequence. When the user issues a command to
start tracking, the camera captures these four consecutive frames and
tries to detect the ArUco marker position by binarizing the marker
frame with the pixel-wise threshold given by the average intensity of
the white and black frames. If the detection fails, the next 
four consecutive frames are captured and the same procedure is repeated.

Once detected, the four corners of the projected ArUco marker are used
to determine the surface area onto which the projected content is
mapped. In our implementation, four corners of the quadrangle area
are used as control points to specify the area, which are called the
{\it surface corners} in the rest of the paper. Note that these are
only imaginary points acting as anchors for control and need
not correspond to feature-rich points in the surface texture.
The template images $I^{+}$ and $I^{-}$ are then sampled from this
area.  

\subsection{Tracking Control}

During the tracking phase, after capturing each newest camera frame, the
tracking procedure described in \autoref{sec:fiducial_track} is
executed using the surface corners in the previous frame applied as the
initial values for the optimization. This procedure outputs the tracking
results of the surface corners and the {\it fiducial corners}, which
are defined as control points in the fiducial pattern that should
coincide with the surface corners.

From the tracked fiducial corner coordinates in the camera image space
and the known fiducial corner coordinates in the projector image
space, we obtain a homography matrix $H_\text{pc}$ with respect to
the surface plane to map a camera image point to its corresponding
projector image point. By mapping the surface corners into the
projector image space using this homography matrix, we have the four
goal points toward which the fiducial corners are regulated in the
projector image space. 

For this control, Kagami and Hashimoto~\cite{kagami2015} applied a
simple proportional-derivative (PD) controller to each of these four
corners.
Although they demonstrated that this achieves
a fairly good tracking performance when a solid-color surface is used as
a target, small tracking errors caused by the feedback-only control
can cause a perceptible misalignment when a textured target surface is
used. Therefore, we introduce a simple implementation of the Smith
predictor. In general, a control system using the Smith predictor for a
discrete time system is given by \cite{warwick1988}:
\begin{align*}
u(z) &= C(z) e(z)\\
e(z) &= y_\text{d} - y + (z^{-k} - 1) G(z) u(z)
\end{align*}
where $u$ is the output of the controller $C$, $e$ is the error input to the
controller, $G$ is the plant model, $y$ and $y_\text{d}$ are the
observed (i.e.\ fiducial corner position) and desired 
(i.e.\ surface corner position) outputs, respectively, and $z^{-k}$ is the
delay operator with dead time $k$ [camera frames]. Using a PD controller
$C(z) = K_\text{p} + K_\text{d} (1 - z^{-1}) $ which takes the position
error as input and generates the velocity output of a corner point, a
simplest plant model is the pure integrator $G(z) = \sum_{n =
  1}^\infty z^{-n}$, meaning that the unknown motion of the target
surface is simply neglected. Plugging these definitions into the above
equations yields
\begin{align*}
u(z) &= \{ K_\text{p} + K_\text{d} (1 - z^{-1}) \} e(z)\\
e(z) &= y_\text{d} - y - \sum_{n = 1}^{k} z^{-n} u(z). 
\end{align*}
This means that we only need to memorize the recent $k$ outputs from
the PD controller and to add their sum to the observed fiducial
corner position $y$.

In our implementation, dead time $k = 1$ was chosen because even an
extremely small $K_\text{p}$ with $K_\text{d} = 0$ cannot stabilize
the system when the dead time is set to $k = 2$. This suggests that the
overall system latency is longer than 1 camera frame and shorter than
2 camera frames (i.e.,\ between 2.5 and 5.0~ms). The controller
gains were empirically chosen to be $K_\text{p} = 0.15$ and $K_\text{d}
= 4.0$.

Once the values of $u$ are computed for the four corners, they are added to the
current fiducial corner positions in the projector image space to
determine the next fiducial corner positions. The homography matrix
for the projected content is computed from these positions and sent to
the projector.

\section{Quantitative Evaluation}

\begin{figure}[tb]
\centering
\includegraphics[width=0.95\columnwidth]{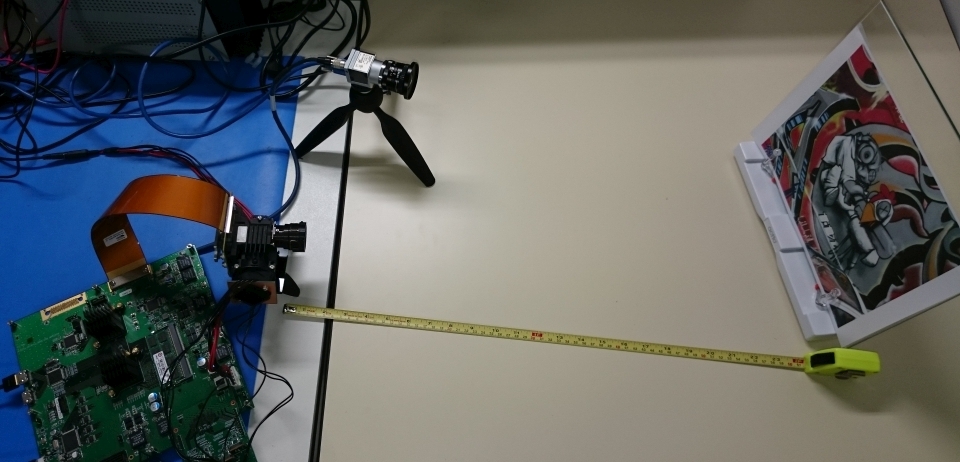}
\caption{Overview of evaluation setup.}
\label{fig:eval_setup}
\end{figure}

\begin{figure}[tb]
\centering
\includegraphics[width=0.9\columnwidth]{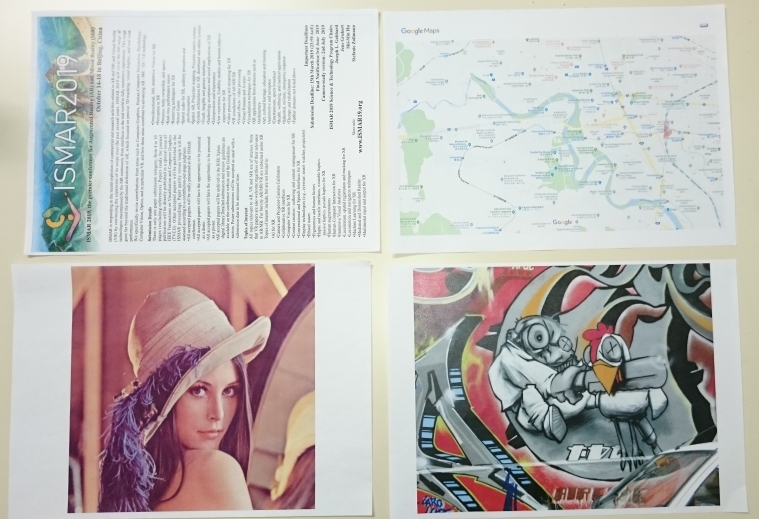}
\caption{Target surfaces used for evaluation.}
\label{fig:eval_surfaces}
\end{figure}

\begin{figure}[tb]
\centering
\includegraphics[width=0.45\columnwidth]{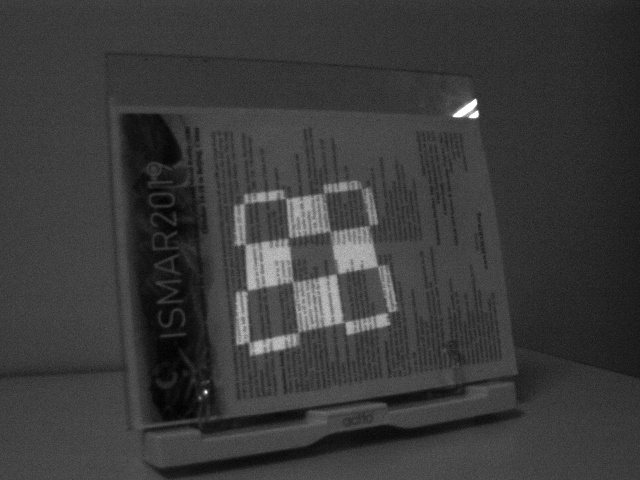}~%
\includegraphics[width=0.45\columnwidth]{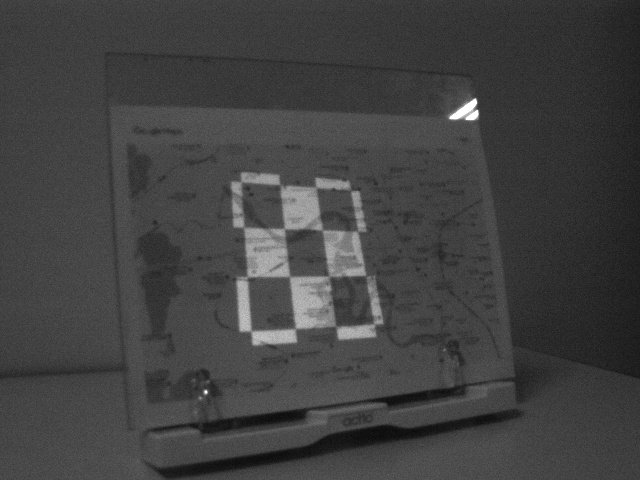}\\[1mm]
\includegraphics[width=0.45\columnwidth]{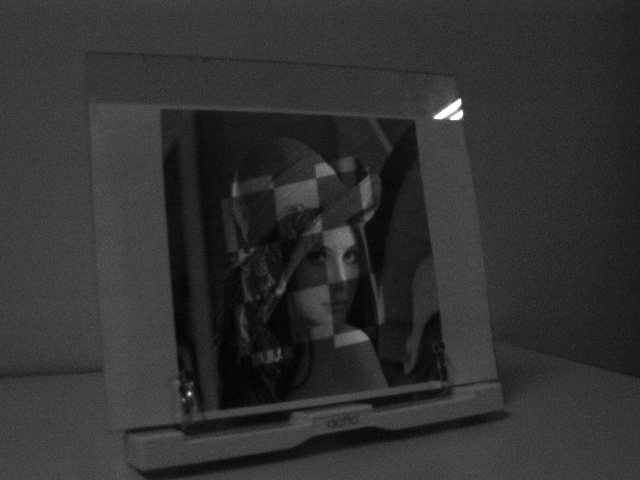}~%
\includegraphics[width=0.45\columnwidth]{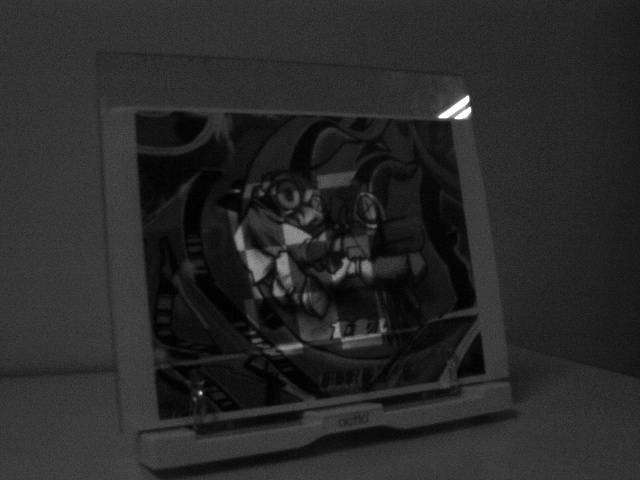}
\caption{Examples of captured images in the evaluation test.}
\label{fig:eval_cam_image}
\end{figure}

\begin{figure}[tb]
\centering
\includegraphics[width=0.9\columnwidth]{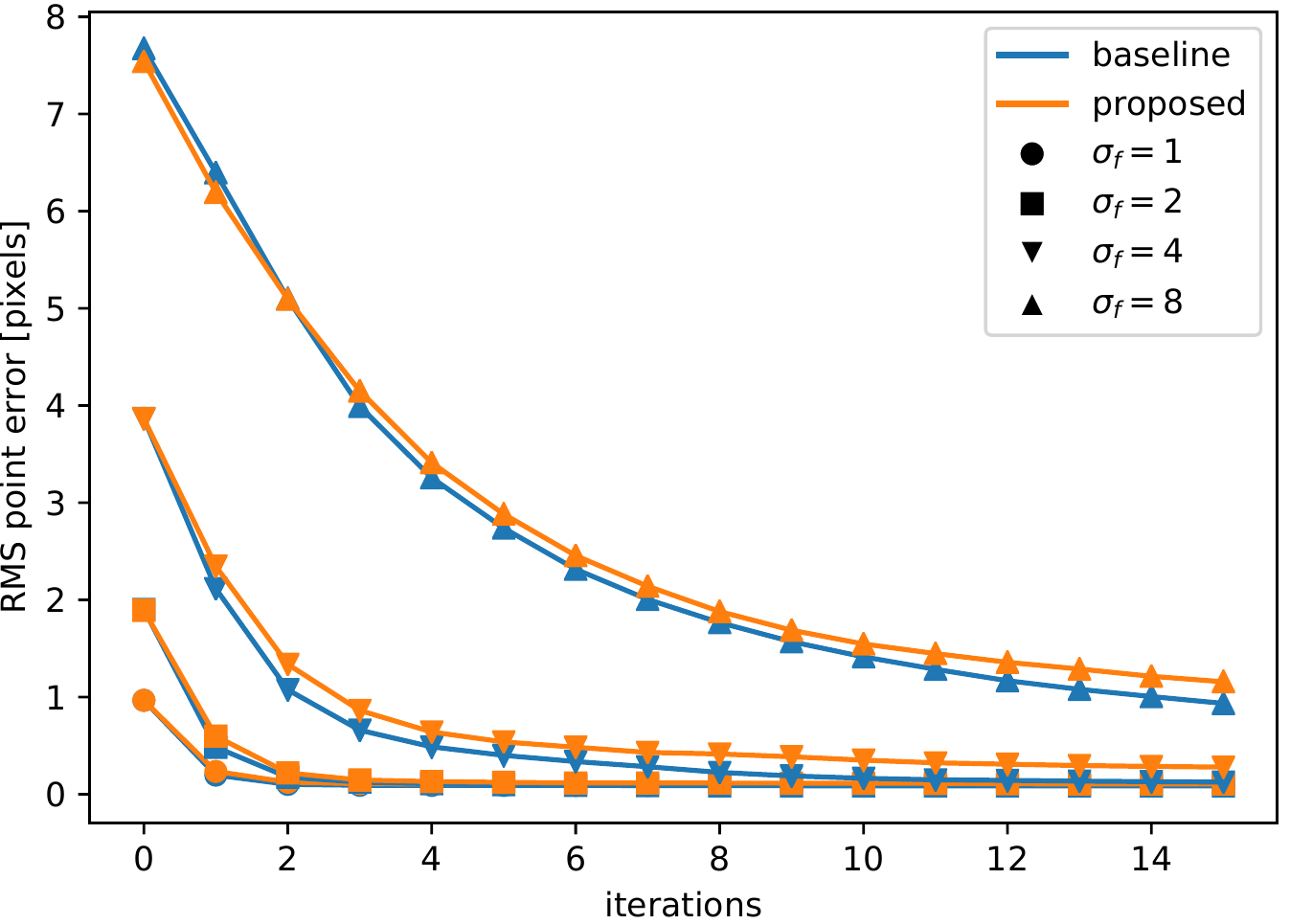}
\caption{Average RMS point errors of surface tracking versus iteration count when no
  projection interference takes place.}
\label{fig:rmse_surface_allon}
\end{figure}

\begin{figure}[tb]
\centering
\includegraphics[width=0.9\columnwidth]{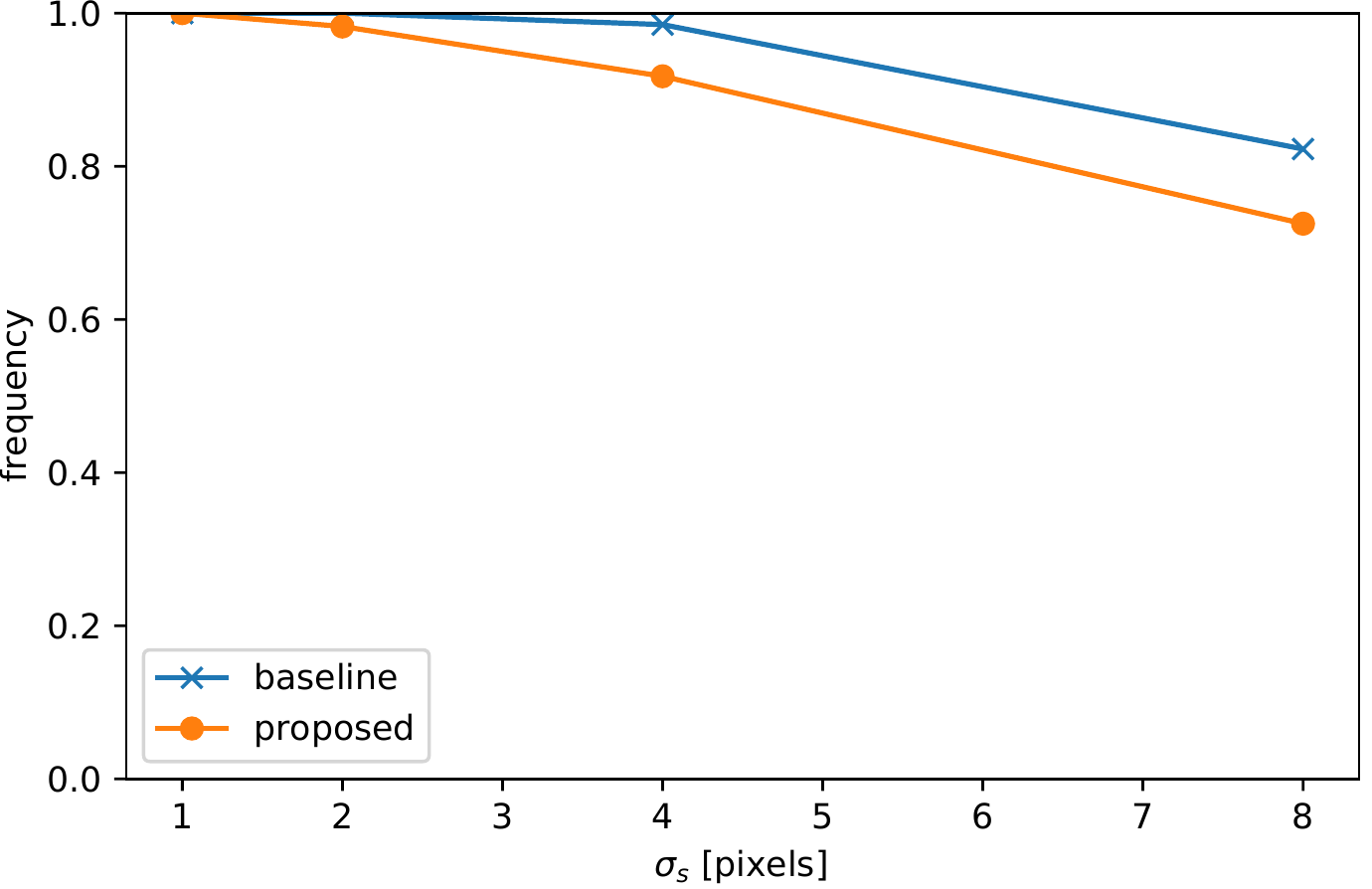}
\caption{Frequency of convergence of surface tracking when no
  projection interference takes place.}
\label{fig:cfreq_surface_allon}
\end{figure}

\begin{figure}[p]
\centering
\includegraphics[width=0.9\columnwidth]{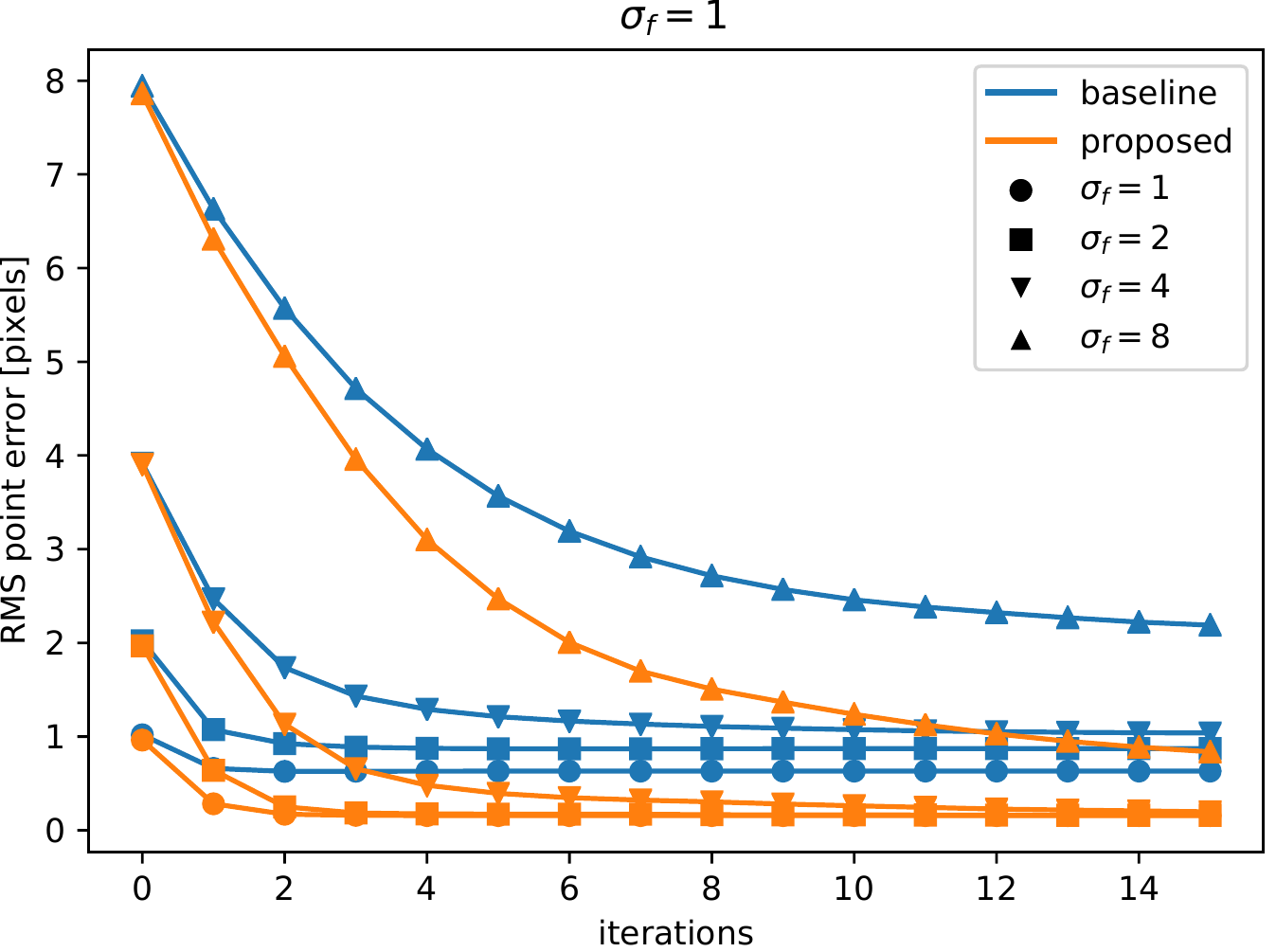}\\[-2.4mm]
\includegraphics[width=0.9\columnwidth]{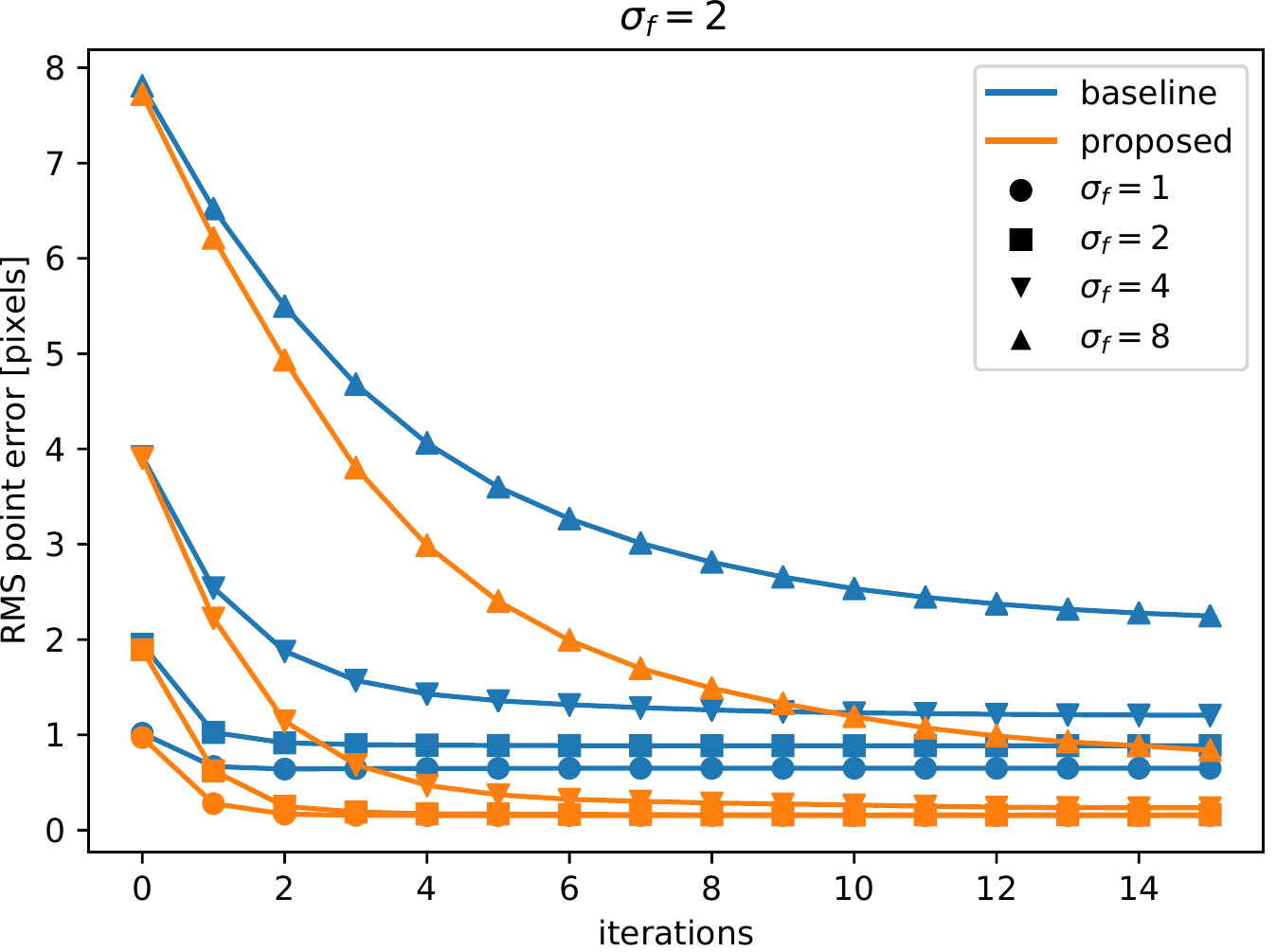}\\[-2.4mm]
\includegraphics[width=0.9\columnwidth]{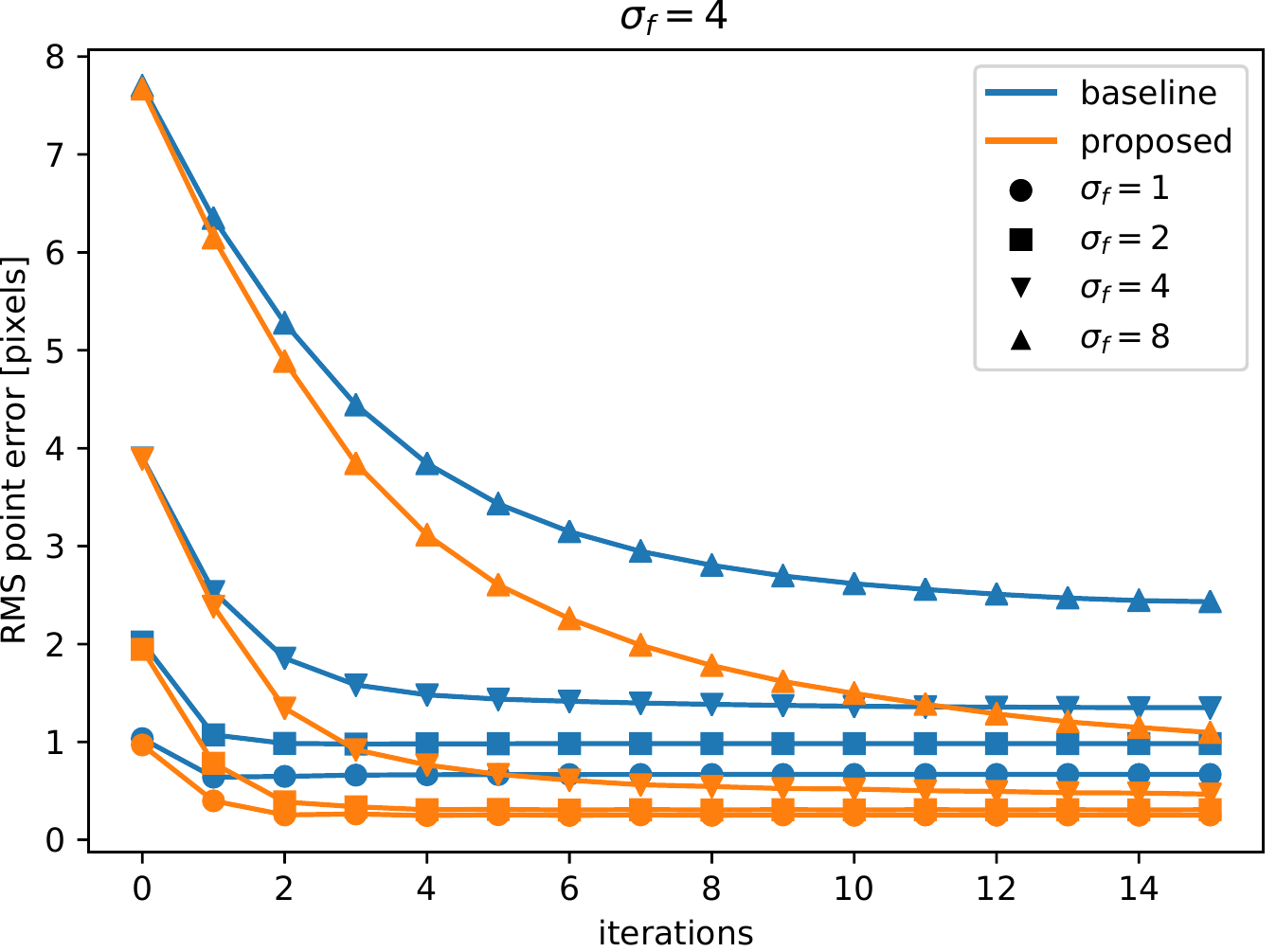}\\[-2.4mm]
\includegraphics[width=0.9\columnwidth]{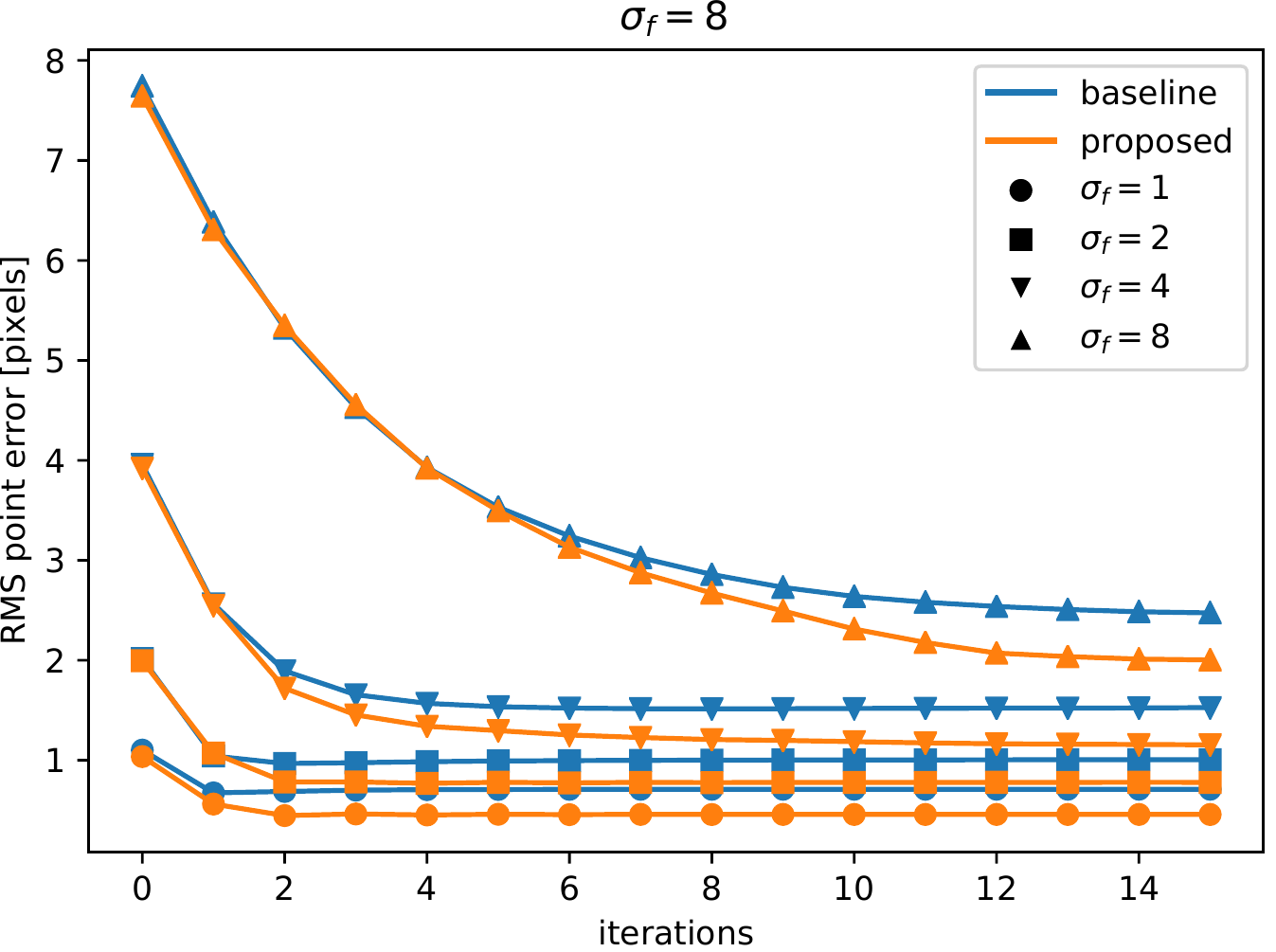}
\caption{Average RMS point errors of surface tracking versus number of iterations when interference from a fiducial projection occurs.}
\label{fig:rmse_surface}
\end{figure}

\begin{figure}[p]
\centering
\includegraphics[width=0.9\columnwidth]{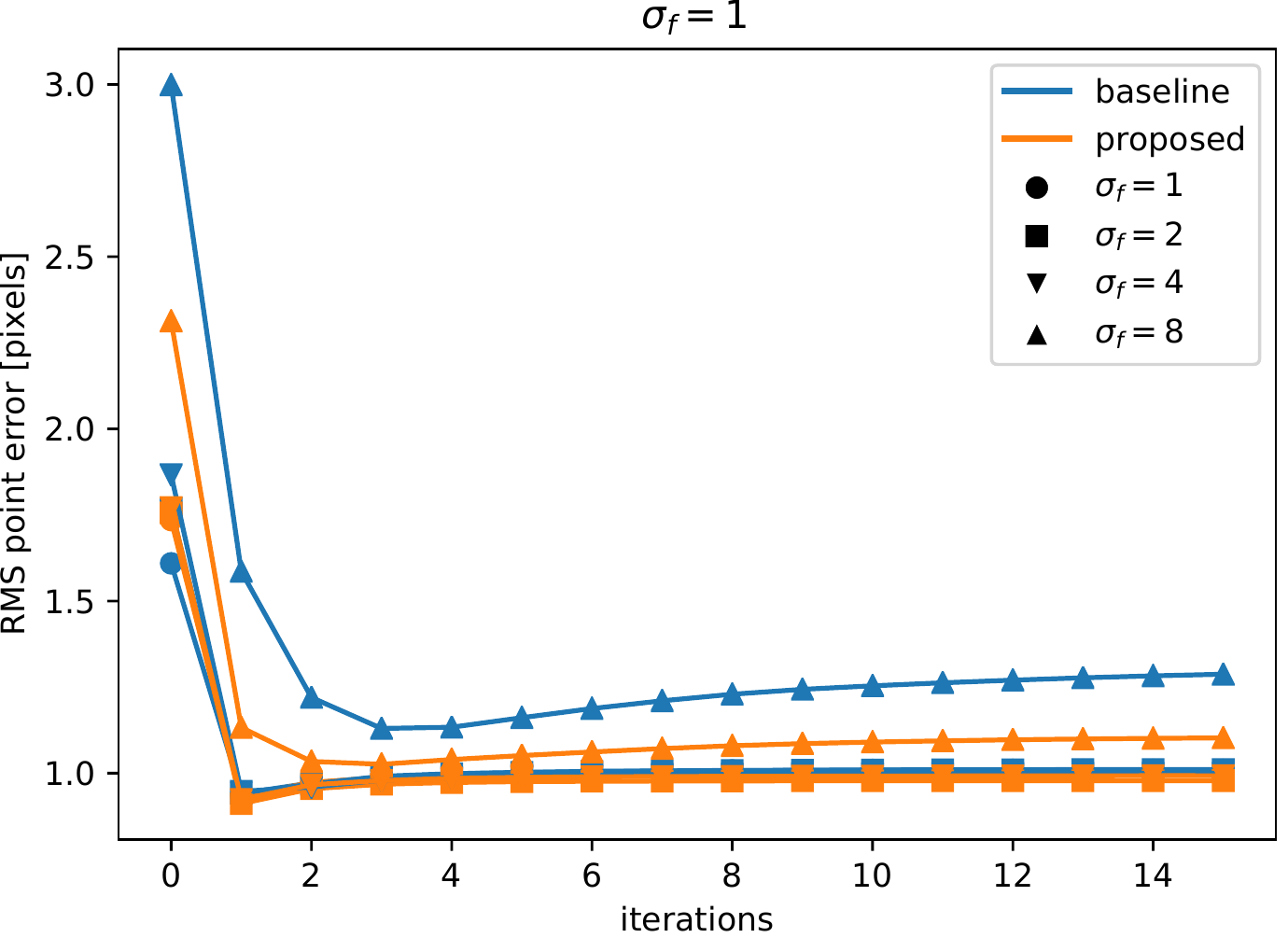}\\[-2.4mm]
\includegraphics[width=0.9\columnwidth]{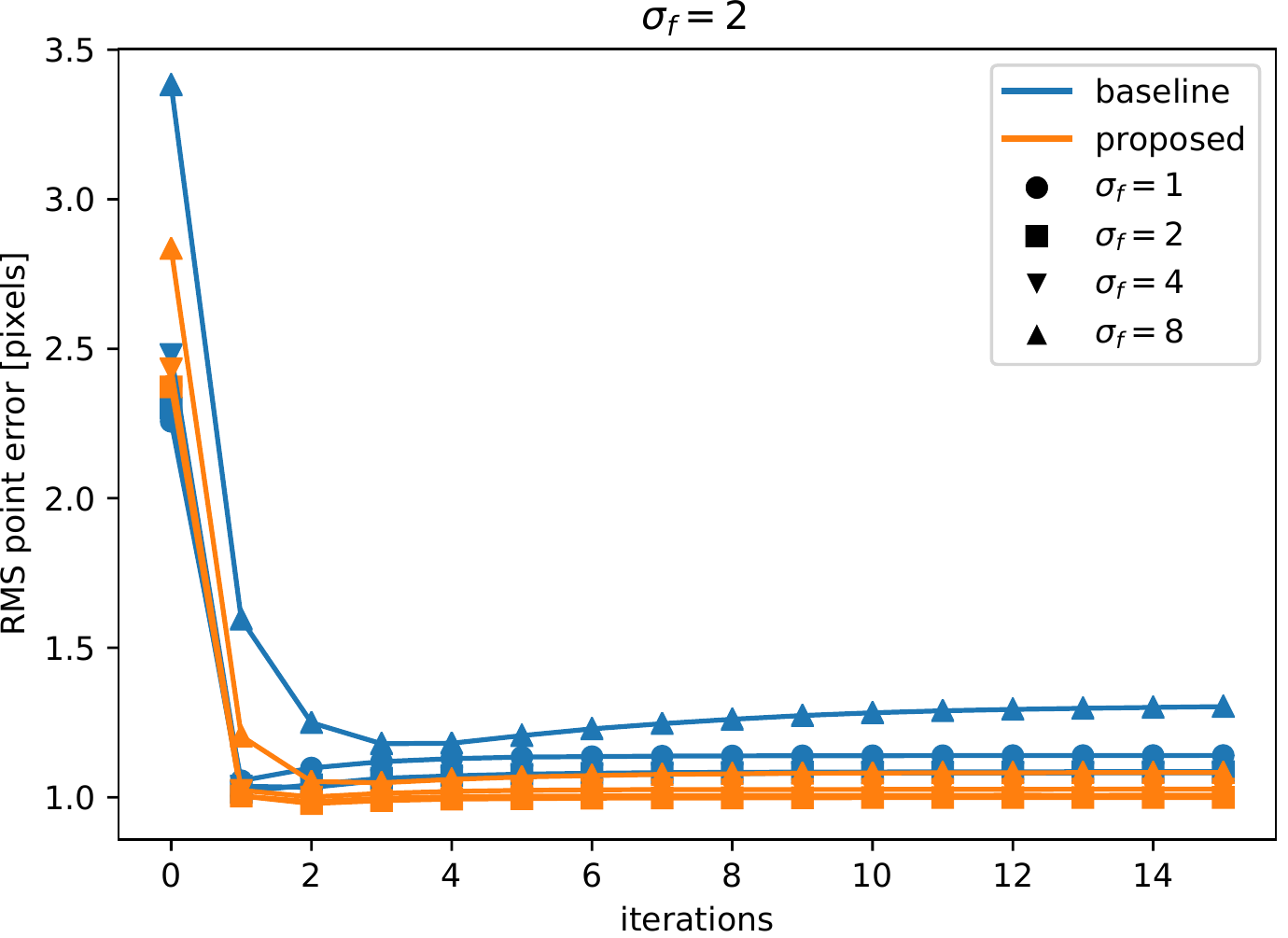}\\[-2.4mm]
\includegraphics[width=0.9\columnwidth]{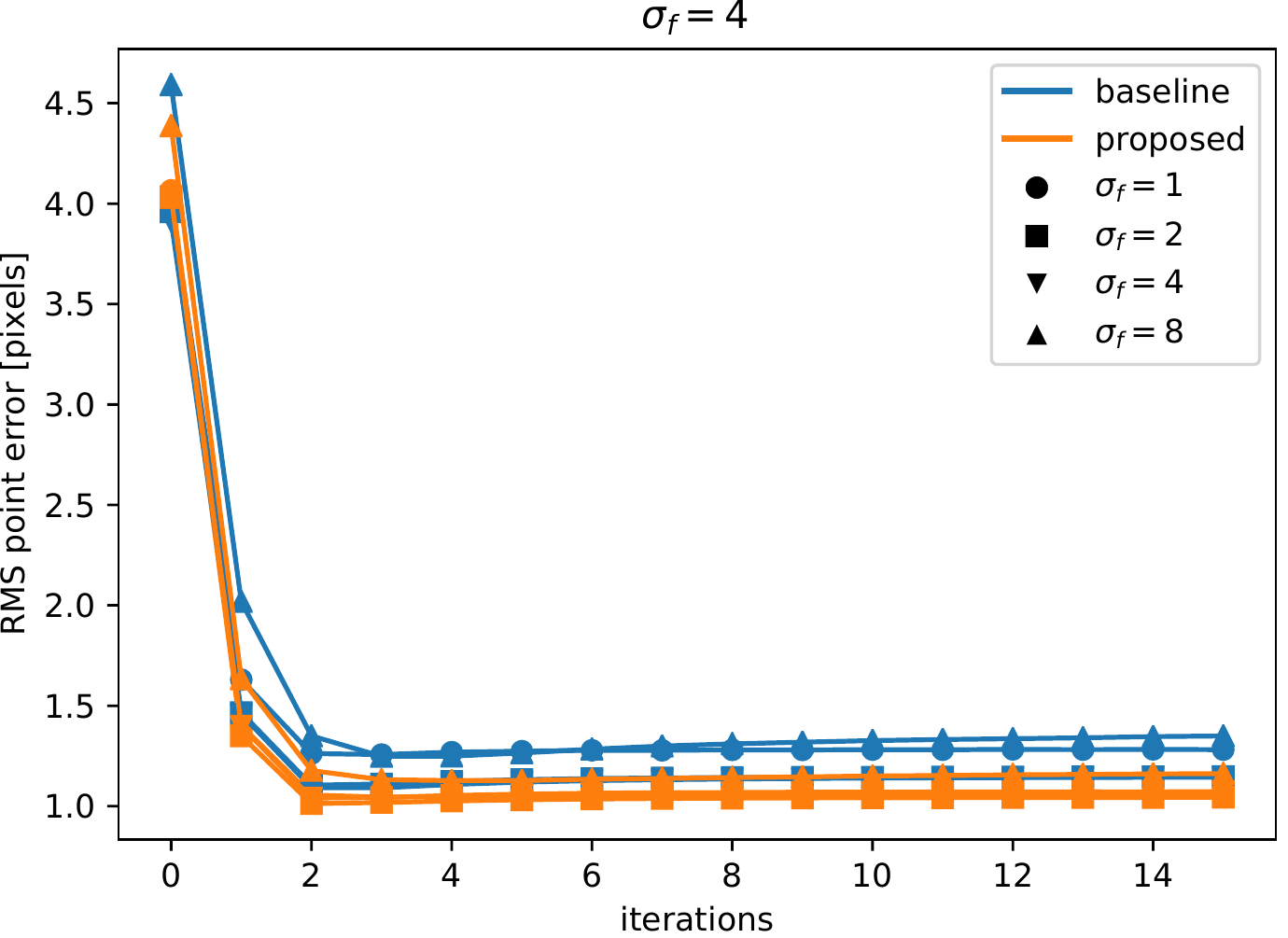}\\[-2.4mm]
\includegraphics[width=0.9\columnwidth]{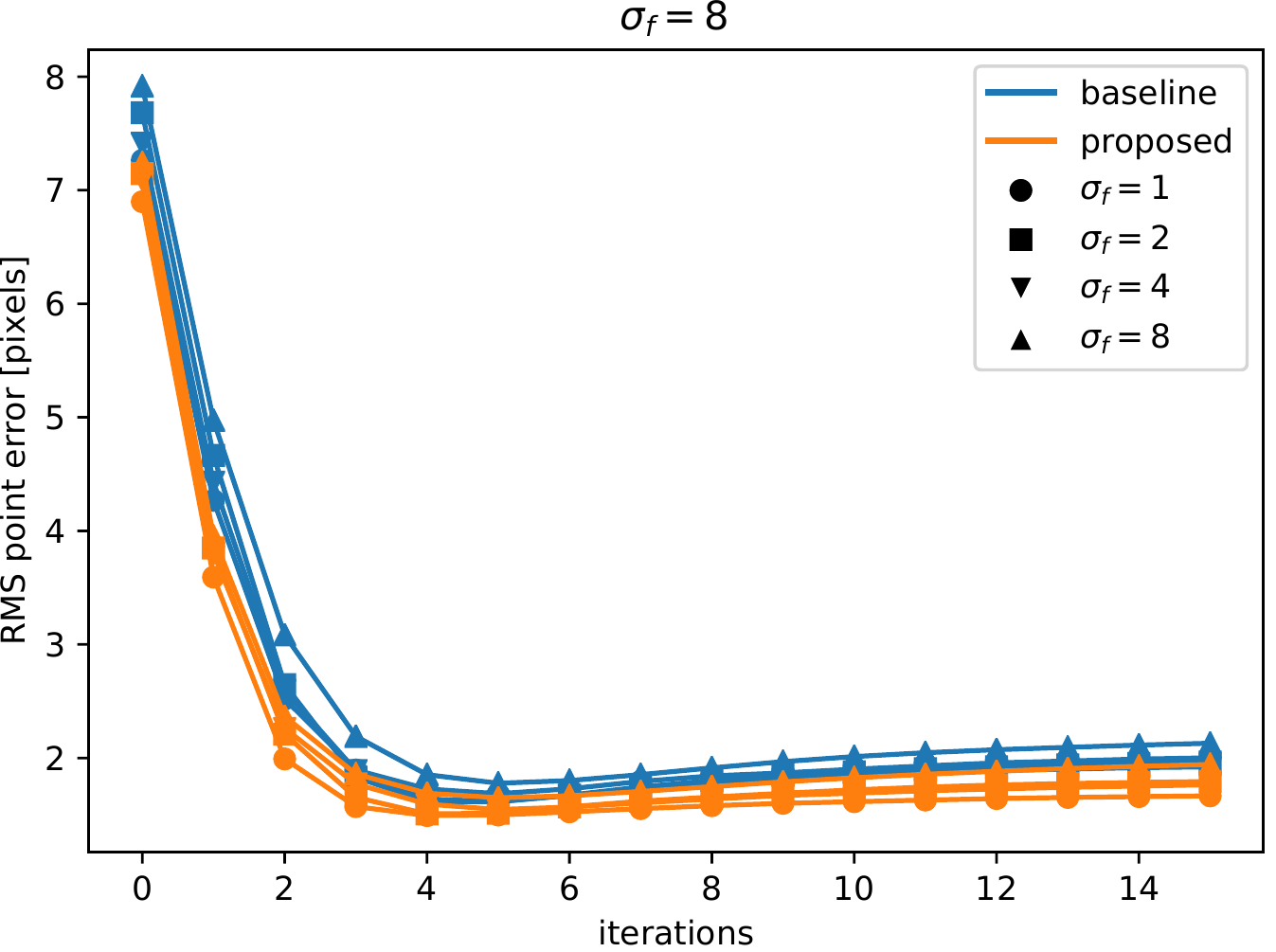}
\caption{Average RMS point errors of fiducial tracking versus number of iterations when interference from a fiducial projection occurs.}
\label{fig:rmse_fiducial}
\end{figure}

\begin{figure}[tb]
\centering
\includegraphics[width=0.9\columnwidth]{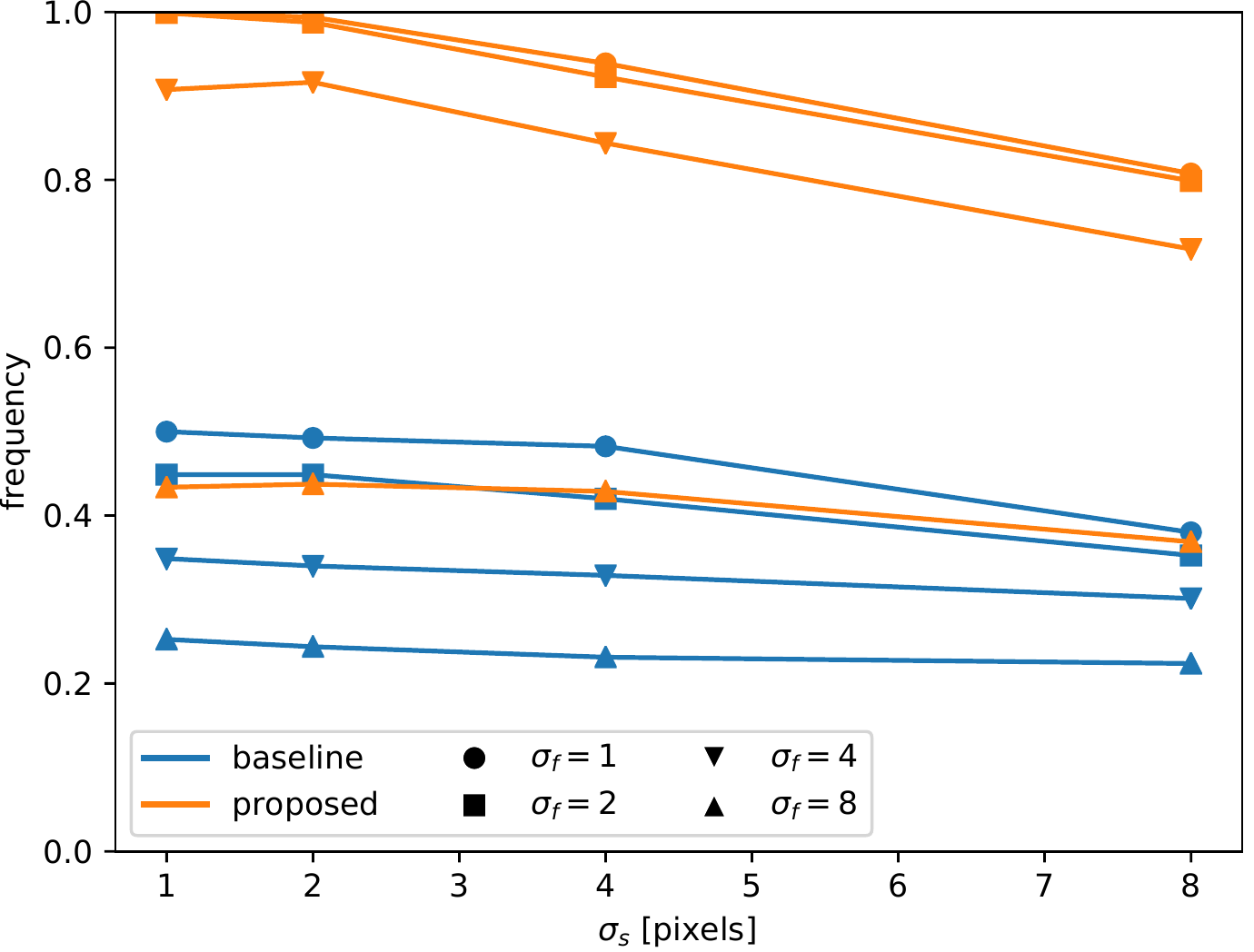}
\caption{Frequency of convergence of surface tracking.}
\label{fig:cfreq_surface}
\end{figure}

\begin{figure}[tb]
\centering
\includegraphics[width=0.9\columnwidth]{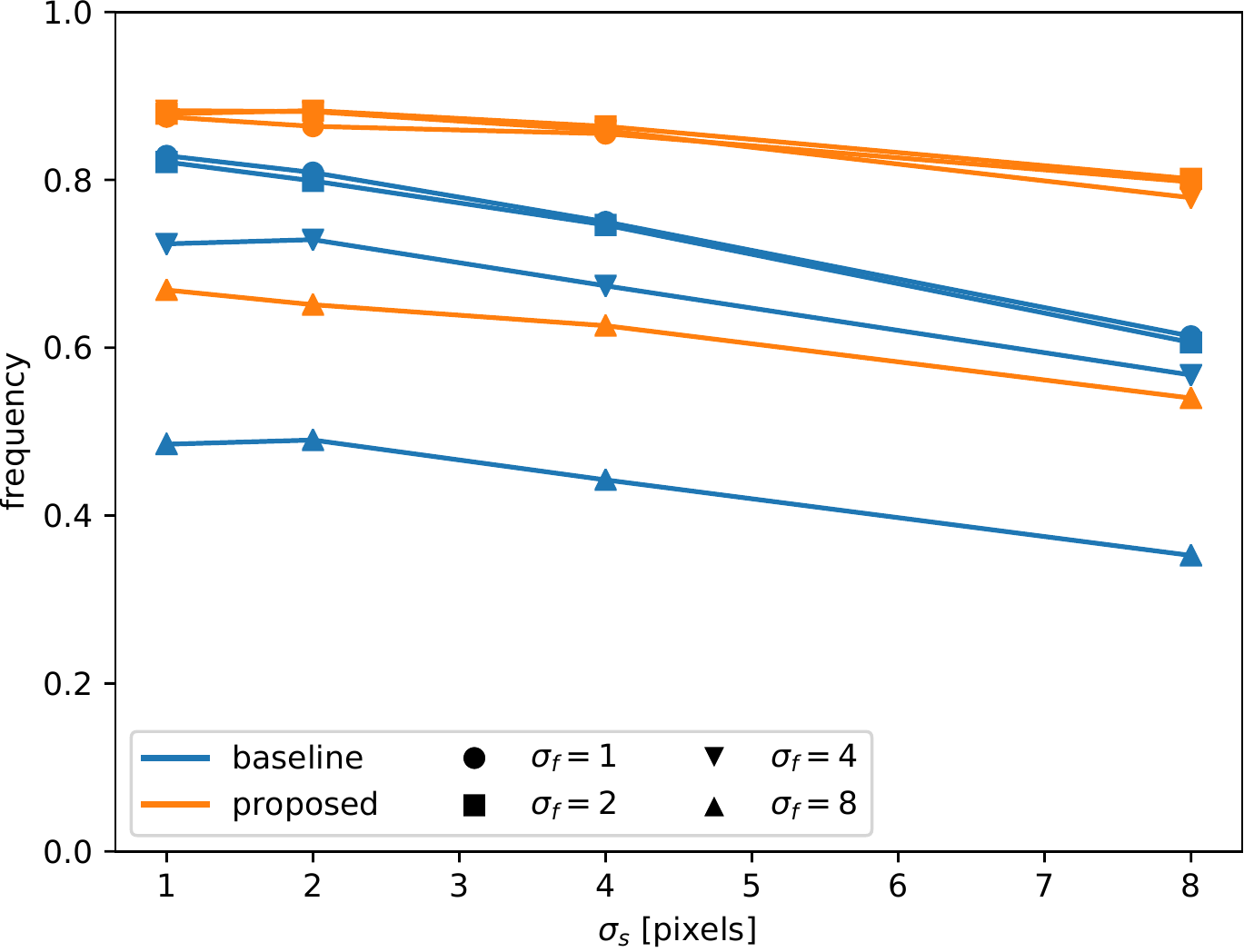}
\caption{Frequency of convergence of fiducial tracking.}
\label{fig:cfreq_fiducial}
\end{figure}

\begin{figure}[tb]
\centering
\includegraphics[width=0.9\columnwidth]{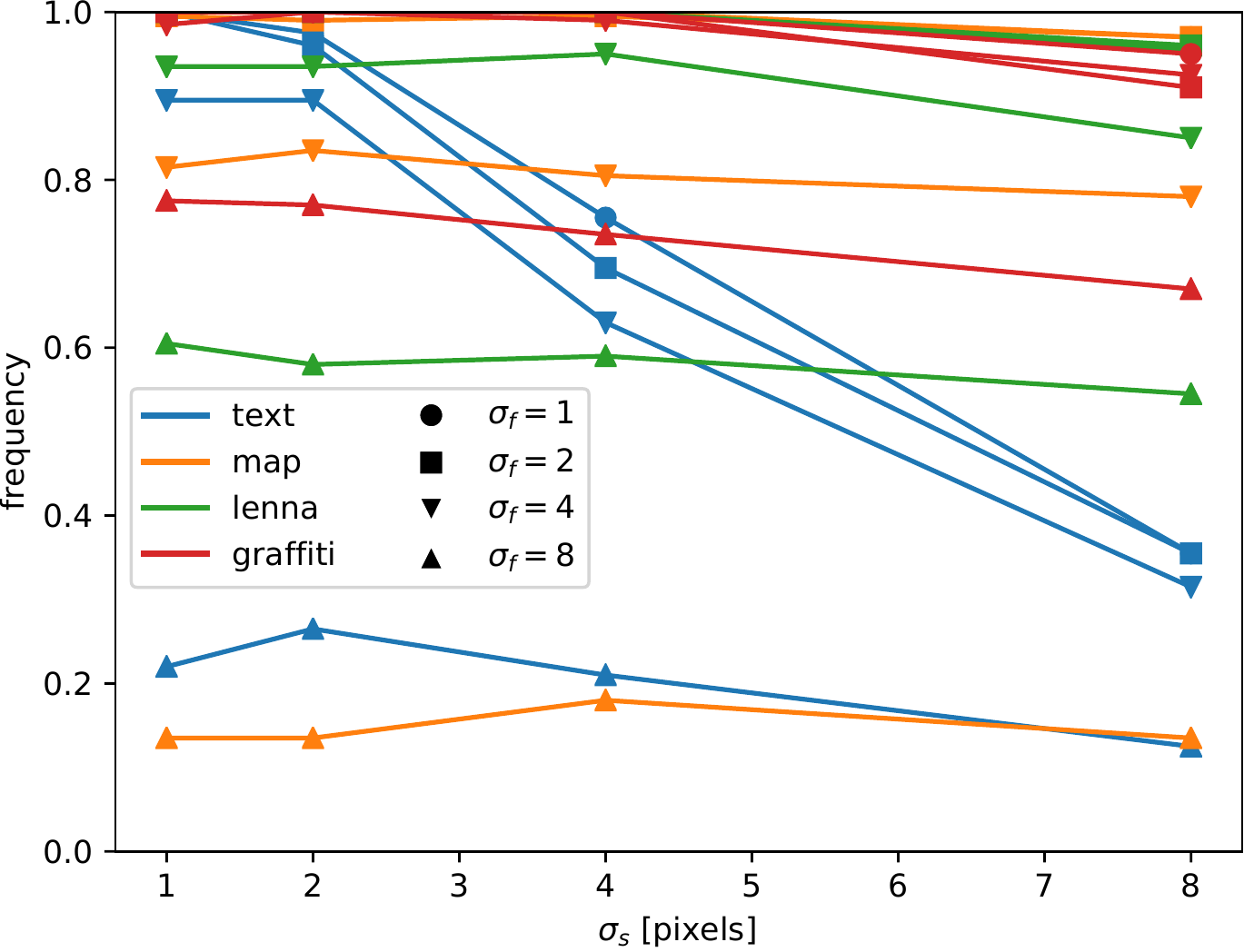}
\caption{Frequency of convergence of surface tracking with respect
  to different surface textures.}
\label{fig:cfreq_surface_for_surftex}
\end{figure}

\begin{figure}[tb]
\centering
\includegraphics[width=0.9\columnwidth]{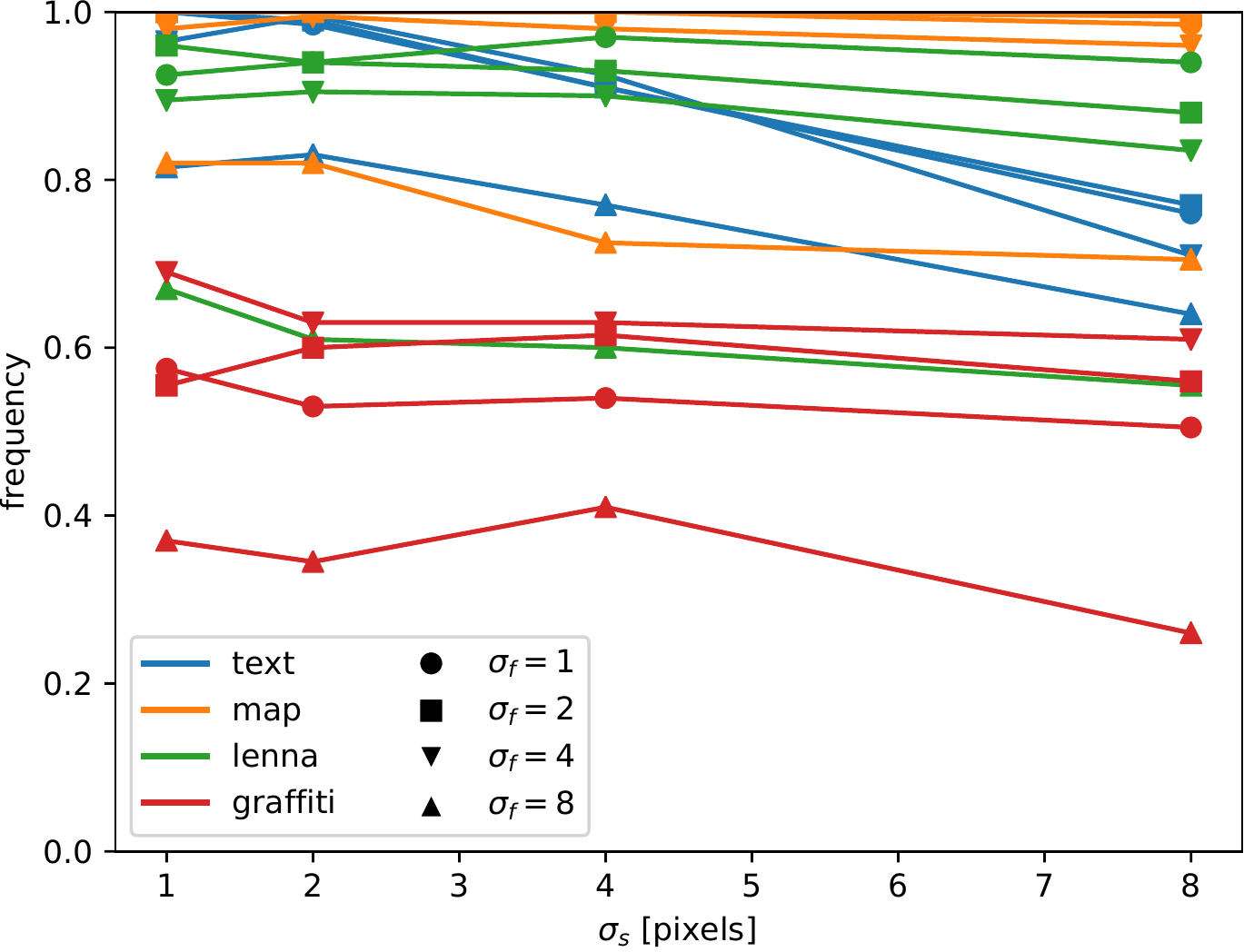}
\caption{Frequency of convergence of fiducial tracking with respect
  to different surface textures.}
\label{fig:cfreq_fiducial_for_surftex}
\end{figure}

Our goal is to achieve a mapping of the projected contents onto a fast and
randomly moving surface; however, it is difficult to evaluate the
performance of such a system quantitatively, mainly because the
accurate ground truth is difficult to obtain. However, a
simulation using synthesized images does not consider the
various sources of a disturbance.

We therefore decided to prepare many real images of fixed surfaces onto
which randomly warped fiducial patterns are projected, and run the
tracking algorithm from randomly deviated initial surface corner
points from the ground truth to see if the results converge with the
ground truth. 

\subsection{Dataset Acquisition}

We ran the proposed method for a fixed surface set in front of the
projector-camera system and made sure that it converged to seemingly
correct positions through an eye observation. The surface corner positions
tracked in the camera image space at this time instant were stored as
the {\it ground truth surface corner points}, which are common for
all the images taken with this surface. 

At the same time, the fiducial corner positions tracked in the camera
image space and the known fiducial corners in the projector image
space were used to estimate the homography matrix
$\tilde{H}_\text{pc}$ with respect to this surface.

To generate randomly warped fiducial patterns, random numbers
drawn from a normal distribution $N(0, \sigma_\text{f}^2)$ were added to
the ground truth surface corners to generate the {\it ground truth fiducial corner
points}. These corner points were mapped by $\tilde{H}_\text{pc}$ to
the projector image space to control the projector such that the corners
of the projected fiducial patterns were observed in the camera image at
the above-defined ground truth fiducial corners. The camera images of
this scene were captured 100 times for each of $\sigma_\text{f} =
1, 2, 4, 8$ [pixels] with each of the original fiducial patterns and
their inverted counterparts. In addition, 100 images with all 
projector pixels on, and another set of 100 images with all pixels
off, were captured and stored. This entire procedure was repeated for
four surfaces with different textures: ``Text,'' ``Map,'' ``Lenna,'' and
``Graffiti,'' each of which was printed on a sheet of copier paper using a
color laser printer and stuck onto a glass plate.

\autoref{fig:eval_setup} shows the evaluation setup.  
Images with a pixel resolution of $640 \times 480$ were
taken using a Basler acA640-750um camera with the exposure time set to
400~$\mu$s and the gain set to 16, through a Space HF3.5M-2 C-mount
lens (3.5-mm focal length, F1.6). Illuminance at around the surface
was approximately 210~lx, and the pixel value of a white surface point was
approximately 60 when not illuminated by the projector and approximately 150 when
illuminated by the white light source used to present the fiducials. 
\autoref{fig:eval_surfaces} shows the four surfaces used for
the evaluation and \autoref{fig:eval_cam_image} presents examples of
the captured images. 

\subsection{Tracking Test Procedure}

Using the above dataset, off-line optimization tests were carried out
with randomly deviated initial values to see the performance with
respect to the convergence rate of optimization and how frequently
the optimization converged. Random numbers drawn from $N(0,
\sigma_\text{s}^2)$ were added to the ground truth surface corners to
generate the initial corner points from which optimization begins.

We carried out 100 trials of optimization for each combination
of the optimization methods, four surface types, two patterns of
fiducials, $\sigma_\text{f} = 1, 2, 4, 8$ [pixels], and
$\sigma_\text{s} = 1, 2, 4, 8$ [pixels]. During each trial, an image with
all projector pixels on, another with all projector pixels off, and
a third with the fiducial pattern retrieved randomly from the
dataset were applied. The template images $I^{+}$ and $I^{-}$ were sampled from the
first two images, respectively, from the quadrangle specified by the
ground-truth surface corners. The other image is used to test the
tracking algorithm from the initial corner positions with
$\sigma_\text{s}$ deviations. For the optimization methods, we applied both the
plain ESM method as the baseline and the proposed method. The number
of iterations for both methods was limited to 15 for surface
and fiducial tracking, respectively.

\subsection{Results}

The convergence rate of the optimization methods in general can be visualized
by plotting the errors versus the number iterations. We evaluated the errors based on the
root mean square (RMS) errors of the four corner point positions from
the ground truth positions. First, we show the results when there
was no projection interference to see the basic performance of the
baseline (plain ESM) and the proposed methods for a common planar
target tracking problem. We used the all-projector-pixels-on images
for tracking during these tests. \autoref{fig:rmse_surface_allon} shows
the evolution of the average RMS point errors with respect to the number of iterations, where the trials that diverged were excluded from the average
as applied by Baker and Matthew \cite{baker2004}. A trial is said to
have diverged when the final RMS error is greater than the initial RMS
error.

Another metric for evaluation is the frequency of convergence, which is
the number of successful trials that have converged divided by the
number of trials. A trial is said to have converged when the
final RMS error is smaller than 1.0 pixel.
\autoref{fig:cfreq_surface_allon} shows the results when no projection
interference occurs.

\autoref{fig:rmse_surface_allon} and \autoref{fig:cfreq_surface_allon}
show that both the baseline and the proposed method work well for
surfaces without interfering light, although the proposed method performs
marginally worse because the number of pixels that
participate in optimization is smaller owing to the pixel selection.

We now move on to the results regarding our primary concern.
\autoref{fig:rmse_surface} shows the average RMS point errors in the
surface tracking versus the number of iterations when there was interference
by a fiducial projection for a different $\sigma_\text{f}$. In contrast
to the case without a projection interference, the plots clearly show
that the proposed method performs better and that the convergence rates of
the proposed method for $\sigma_\text{f} \leq 4$ are comparable to
those without interference. When $\sigma_\text{f}$ is as large as 8,
the performance decreases significantly and the average RMS errors behave
similarly to those of the baseline method. This is because the
assumption of a small inter-frame motion is violated.

\autoref{fig:rmse_fiducial} shows the evolution of the average RMS point
errors in fiducial tracking with respect to the number of
iterations for a different $\sigma_\text{f}$. The averages here include
only those trials in which both the surface tracking and fiducial
tracking did not diverge, because the success of surface tracking is
a prerequisite for that of fiducial tracking. Also note that the baseline
and the proposed methods are the same after the fiducial
tracking starts, namely, both simply execute a plain ESM method. The only
difference between them is the input image, which is affected by the surface tracking result.
The results show that the fiducial tracking converges quickly as a
whole with both methods. Although the results may appear to indicate that
the proposed method performs consistently better, discretion should be applied. Because the projector control used to display the
fiducial patterns was applied using the estimated $\tilde{H}_\text{pc}$,
the ``ground truth'' positions of the fiducial corners possibly
contained small inevitable bias errors. This can be seen from the
fact that the final average RMS error was approximately 1.0 pixel even for the
best performing case. 

\autoref{fig:cfreq_surface} and \autoref{fig:cfreq_fiducial} show
the frequency of convergence of the surface and fiducial
tracking, respectively. Because the final fiducial tracking errors can
be large for the reason described above, 
a trial of the
fiducial tracking is said to have converged if the final RMS error is smaller than
1.5 pixels.  

The results indicate that the frequency of convergence of the proposed
method are almost equally good as long as $\sigma_\text{f} \leq 4$,
although fiducial tracking seems to be more challenging than 
surface tracking.  
In contrast to the convergence rate results of the fiducial tracking,
in which the rates achieved by both methods are comparable, 
the difference between the baseline and the proposed methods is
clearly significant with respect to frequency of convergence of the
fiducial tracking, as well as the surface tracking.

Finally, we examined the influence of different surface textures on the
frequency of convergence. \autoref{fig:cfreq_surface_for_surftex}
shows the results of the surface tracking, whereas
\autoref{fig:cfreq_fiducial_for_surftex} shows those of the fiducial
tracking. These results suggest that the performance depends significantly on
the surface texture. The surface tracking is relatively easy with
the Lenna and Graffiti surfaces. By contrast, Text and a Map are
likely challenging because they contain fine-grained textures for
which large displacements are difficult to deal with using
iterative alignment methods.  
This is particularly the case with a Text surface,
in which the frequency of convergence rapidly decreases with an increase
in $\sigma_\text{s}$. In contrast, for fiducial tracking, Lenna
and Graffiti are rather challenging. This is likely because of the low reflectance
of these surfaces and can be overcome by increasing the brightness
of the fiducial patterns at the sacrifice of the contrast of the video content.

\section{Dynamic Projection Mapping Results}

\begin{figure}[tb]
\centering
\includegraphics[width=\columnwidth]{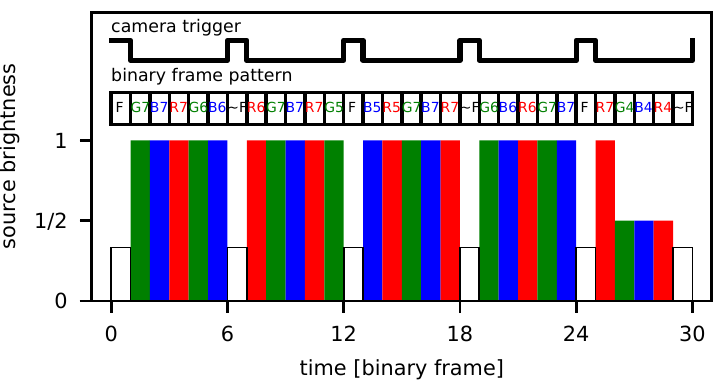}
\caption{A sequence of light source and binary frame switching to
  represent a 4-bit color image with periodically
  inserted fiducial patterns. Binary frames are specified by symbols, such as R7,
  which means the 7th bit (i.e., MSB) of the red channel. Symbols F and
  $\sim$F specify the complementary pair of fiducial patterns.}
\label{fig:seq}
\end{figure}

\begin{figure}[tb]
\centering
\includegraphics[width=\columnwidth]{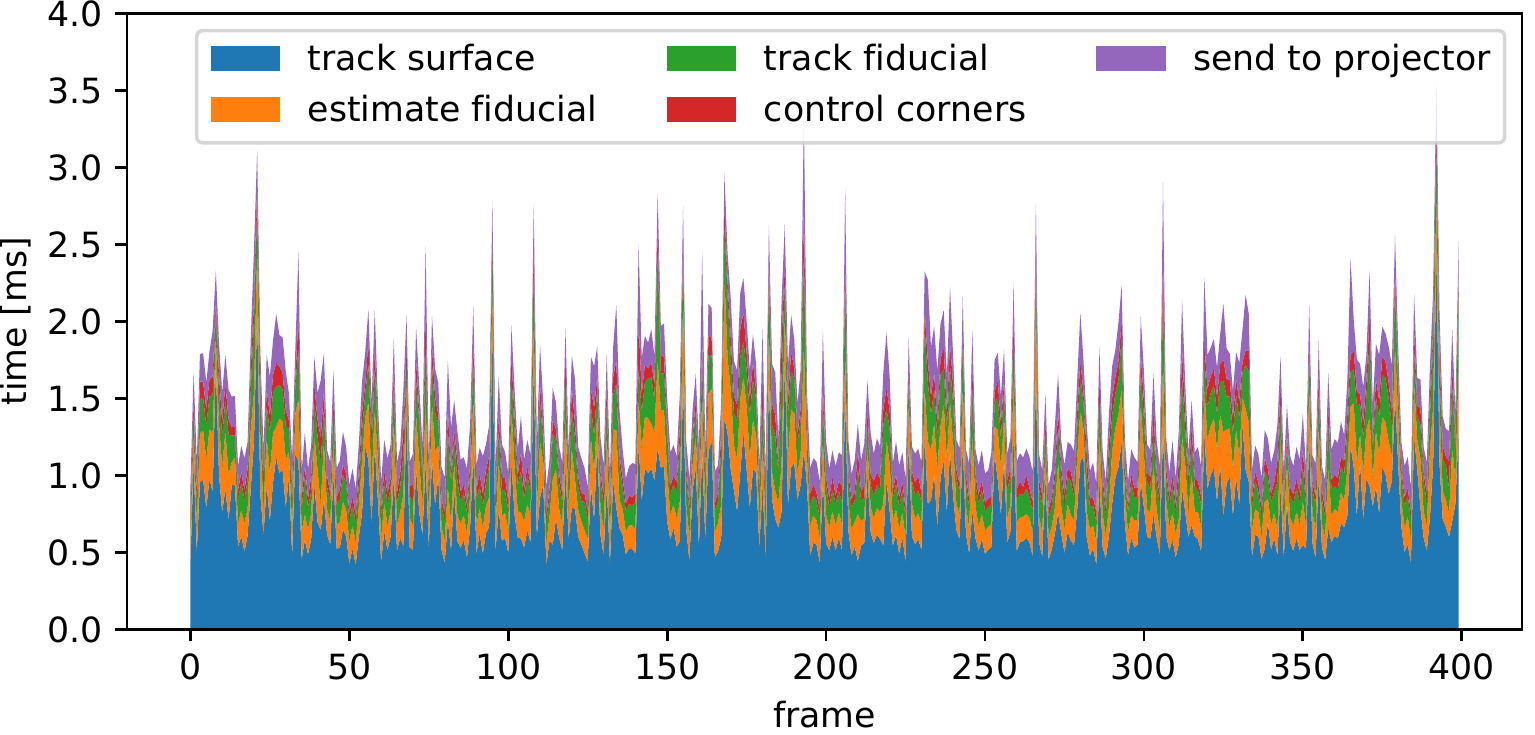}
\caption{Stacked plots of the processing time in a sequence of 400 consecutive camera frames.}
\label{fig:plot_time}
\end{figure}

We implemented the proposed tracking projection algorithm and tested
it for various surfaces with different content. 

To encode a video content with a DMD projector presenting 2,400
binary frames per second, 40 binary frames are available for a video
frame at 60 fps, although we do not need to be so strict about this, and
more or fewer binary frames can participate in the encoding of an image
unless dropped video frames or duplications are fatal for the applications of
interest. Because we reserve $1/6$ of the binary frames for the
fiducial patterns, approximately 33 binary frames are available to represent a
video content. This is sufficient for representing an 8-bit full color
image if we intensively utilize the light source modulation technique,
for which the availability of 24 binary frames is the minimal requirement.
However, herein we employ a 4-bit color representation in favor of better
light utilization. The employed binary frame sequence in our
experiment is shown in \autoref{fig:seq}, although the choice is not
limited to this.

To achieve real-time tracking at 400~fps, we used a
parallelized implementation \cite{kagami2016} of the ESM algorithm using
Intel AVX instructions and multithreading with OpenMP. We set the
size of the surface template to a pixel resolution of $112 \times 112$ and that of
the fiducial template to $48 \times 48$. The number of iterations was
set to 8 and 5 for the surface and fiducial tracking, respectively.
\autoref{fig:plot_time} shows stacked plots of the processing time of a tracking
projection sequence on a laptop PC with an Intel Core-i7 7600U (2.9~GHz)
and 16 GB of RAM running Microsoft Windows 10 Professional.

\subsection{Tracking Control}

\begin{figure}[tb]
\centering
\begin{minipage}{2.6cm} \centering
\includegraphics[width=2.5cm]{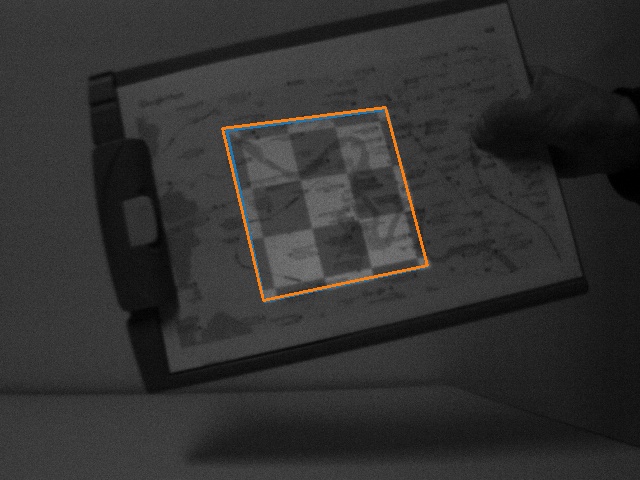}\\
{\footnotesize $t = 0.150 $ [s] }
\end{minipage}~%
\begin{minipage}{2.6cm} \centering
\includegraphics[width=2.5cm]{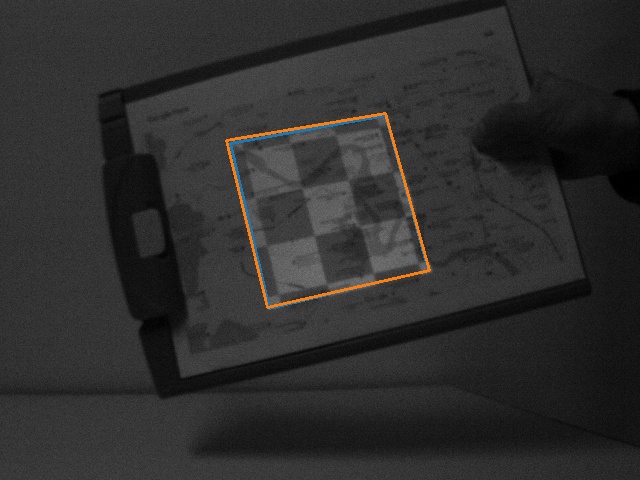}\\
{\footnotesize $t = 0.175 $ [s] }
\end{minipage}~%
\begin{minipage}{2.6cm} \centering
\includegraphics[width=2.5cm]{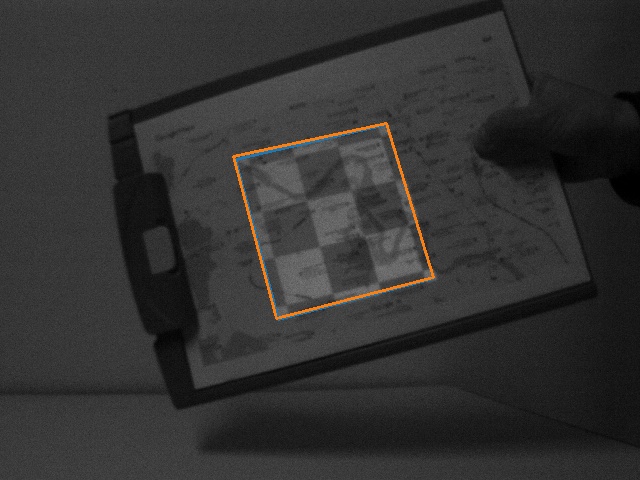}\\
{\footnotesize $t = 0.200 $ [s] }
\end{minipage}\\
\includegraphics[width=\columnwidth]{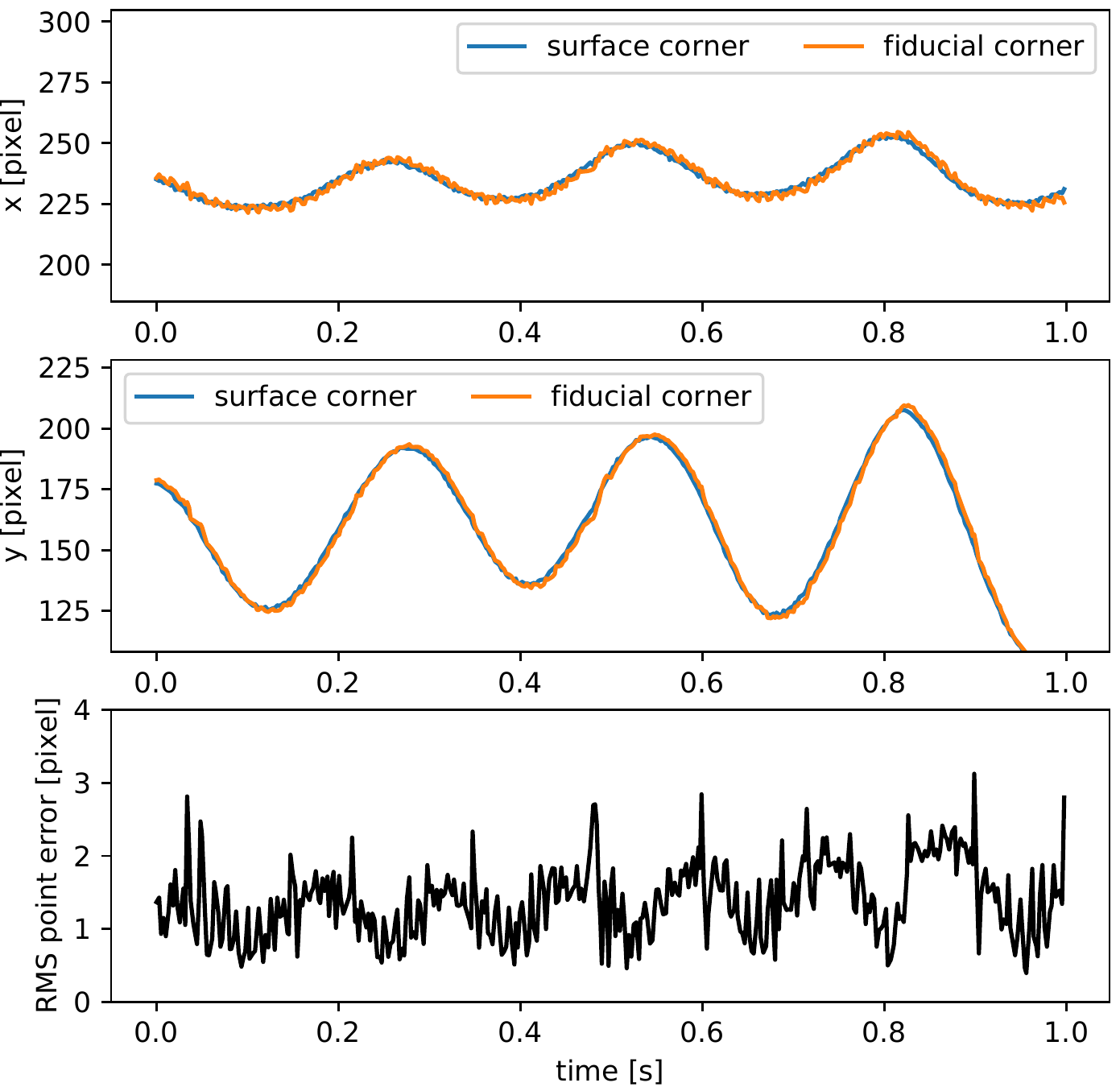}
\caption{The tracking projection results when a surface was in circular
  motion. The top row shows examples of tracking the camera images. The middle
  two rows present the tracked trajectories. The bottom row shows the RMS
  error between the corresponding four corners of the surface and the fiducial
  areas.}
\label{fig:traj_rotate}
\end{figure}

\begin{figure}[tb]
\centering
\begin{minipage}{2.6cm} \centering
\includegraphics[width=2.5cm]{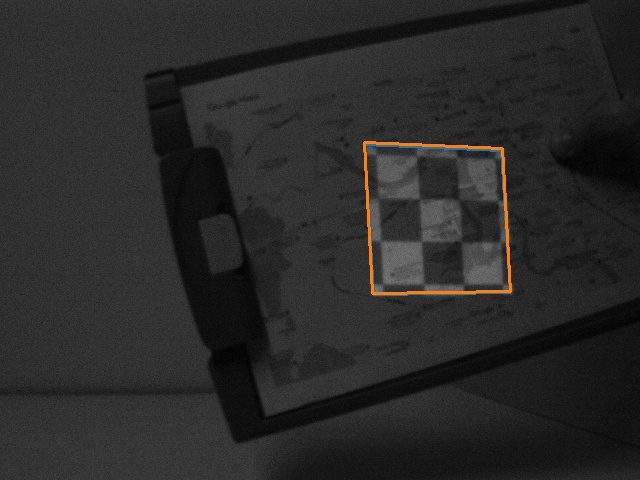}\\
{\footnotesize $t = 0.150 $ [s] }
\end{minipage}~%
\begin{minipage}{2.6cm} \centering
\includegraphics[width=2.5cm]{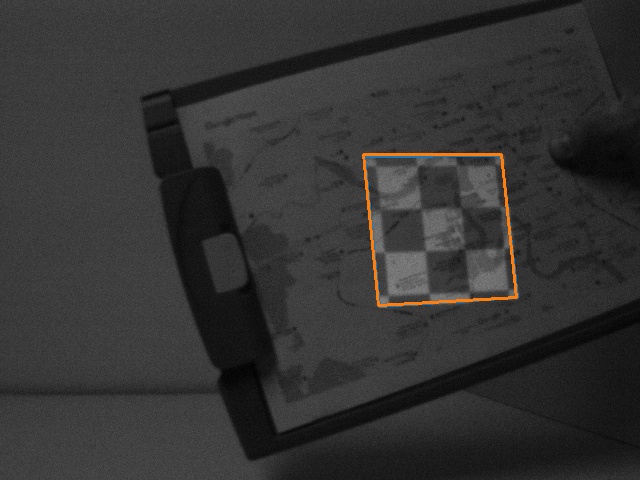}\\
{\footnotesize $t = 0.175 $ [s] }
\end{minipage}~%
\begin{minipage}{2.6cm} \centering
\includegraphics[width=2.5cm]{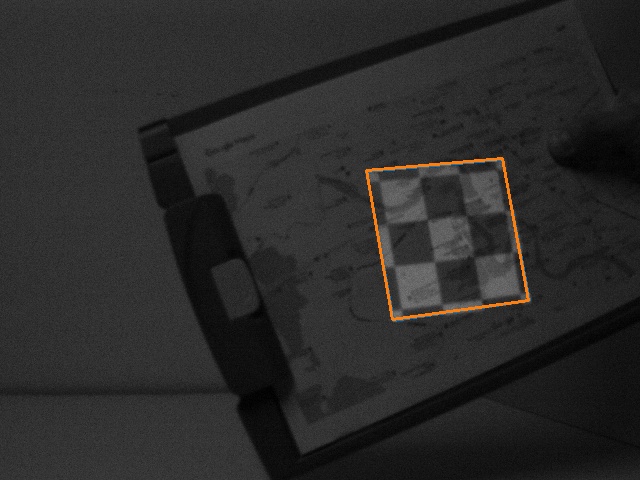}\\
{\footnotesize $t = 0.200 $ [s] }
\end{minipage}\\
\includegraphics[width=\columnwidth]{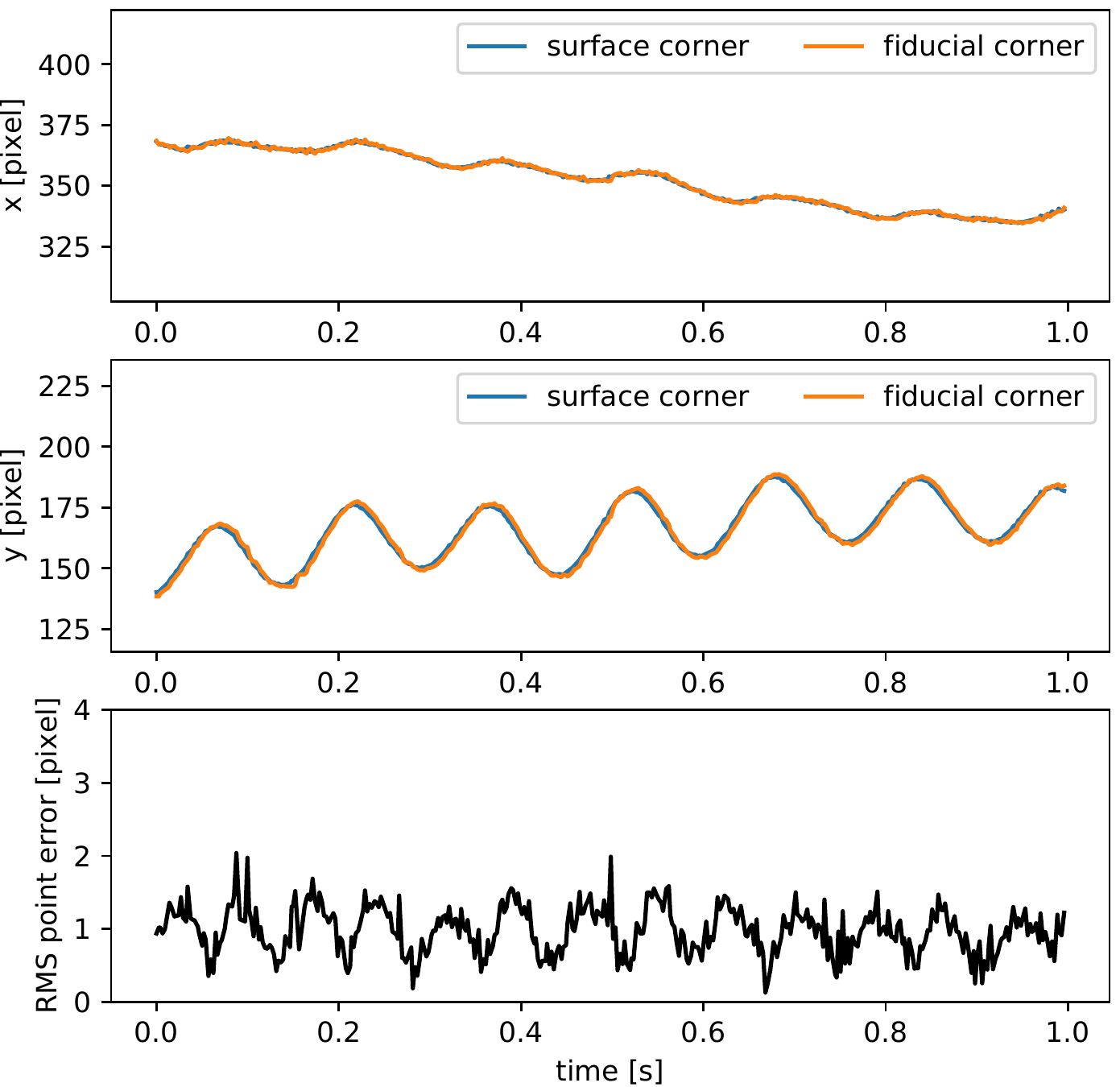}
\caption{Tracking projection results when a surface was vertically shaken.}
\label{fig:traj_shake}
\end{figure}

\begin{figure}[tb]
\centering
\begin{minipage}{2.6cm} \centering
\includegraphics[width=2.5cm]{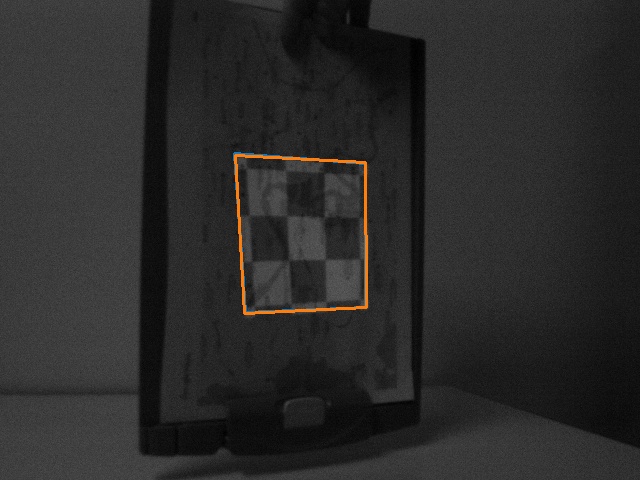}\\
{\footnotesize $t = 0.150 $ [s] }
\end{minipage}~%
\begin{minipage}{2.6cm} \centering
\includegraphics[width=2.5cm]{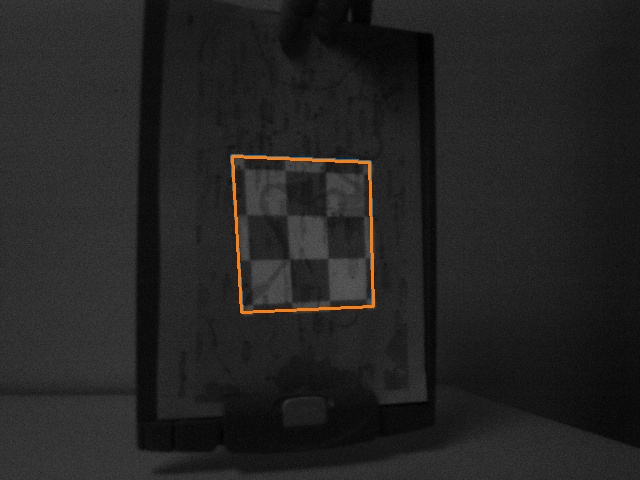}\\
{\footnotesize $t = 0.175 $ [s] }
\end{minipage}~%
\begin{minipage}{2.6cm} \centering
\includegraphics[width=2.5cm]{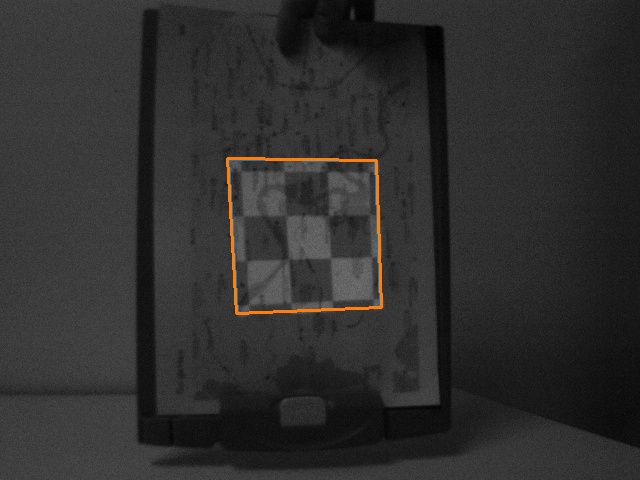}\\
{\footnotesize $t = 0.200 $ [s] }
\end{minipage}\\
\includegraphics[width=\columnwidth]{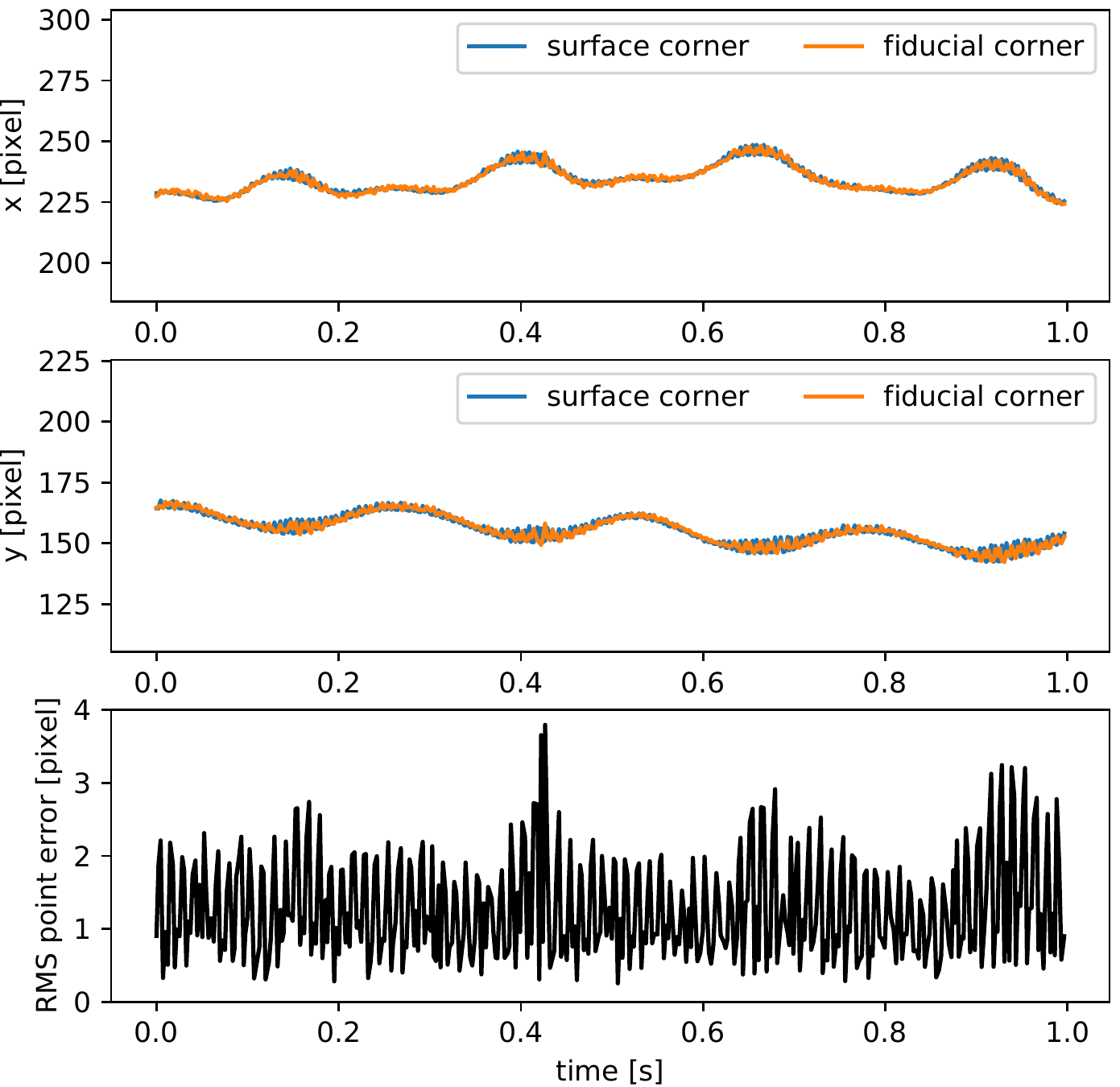}
\caption{Tracking projection results when a surface was spinning about
  the vertical axis.}
\label{fig:traj_yspin}
\end{figure}

\autoref{fig:traj_rotate}, \autoref{fig:traj_shake} and
\autoref{fig:traj_yspin} show the results of the tracking control for
various types of manual surface motions, which include a circular motion
at approximately 4~Hz, vertical shaking at approximately 6~Hz, and spinning about the
vertical axis at approximately 4~Hz, respectively. The $x$ and $y$ plots with
respect to time show trajectories of the upper-left corners of the
surface area and the fiducial in the camera image. The RMS point
errors between the four corresponding pairs of corners are also shown.
Overall, the results indicate that the proposed tracking projection method
works quite well in that RMS errors of the corner positions are within 2
pixels most of the time.

It should be noted that these plots are based on the results of visual
tracking, which themselves may not be accurate, and hence it is
difficult to tell exactly whether the errors are caused by the visual
tracking or the tracking control of the projection. However, it is possible
to conjecture the reasons for particular cases. For example, all three
plots exhibit a periodic variation of the RMS errors corresponding to their
periodic motions. We see larger errors in \autoref{fig:traj_rotate}
and \autoref{fig:traj_shake} when the movement speed is high, and these
seem to have been caused by tracking control errors. In contrast, we
see a constant fluctuation of errors in \autoref{fig:traj_yspin}, and
their marginal increase when the surface was inclined against the
camera. These errors seem to have been caused mainly by visual
tracking errors owing to the challenging conditions rather than the
tracking control.

\subsection{Video Content Mapping}

This subsection presents snapshots of video content mapping
demonstrations. The supplemental video of these scenes shows that the
proposed method enables a fairly quick adaptation of the content onto fast
moving surfaces.

\autoref{fig:cfp} shows snapshots from an external video camera that
captured the scenes of the projection mapping of the video content. The top
row shows a scene in which a rotating logomark of ISMAR2019 was mapped
onto a fast-moving printed CFP. The bottom row shows a similar scene
with an animation video, where the tracking camera on the left hand
side and the surface were being moved by hand, demonstrating
that our approach does not rely on a projector-camera calibration. 

\begin{figure}[tb]
\centering
\subfigure[Video attachment to a piece of paper]{
\includegraphics[width=4.4cm]{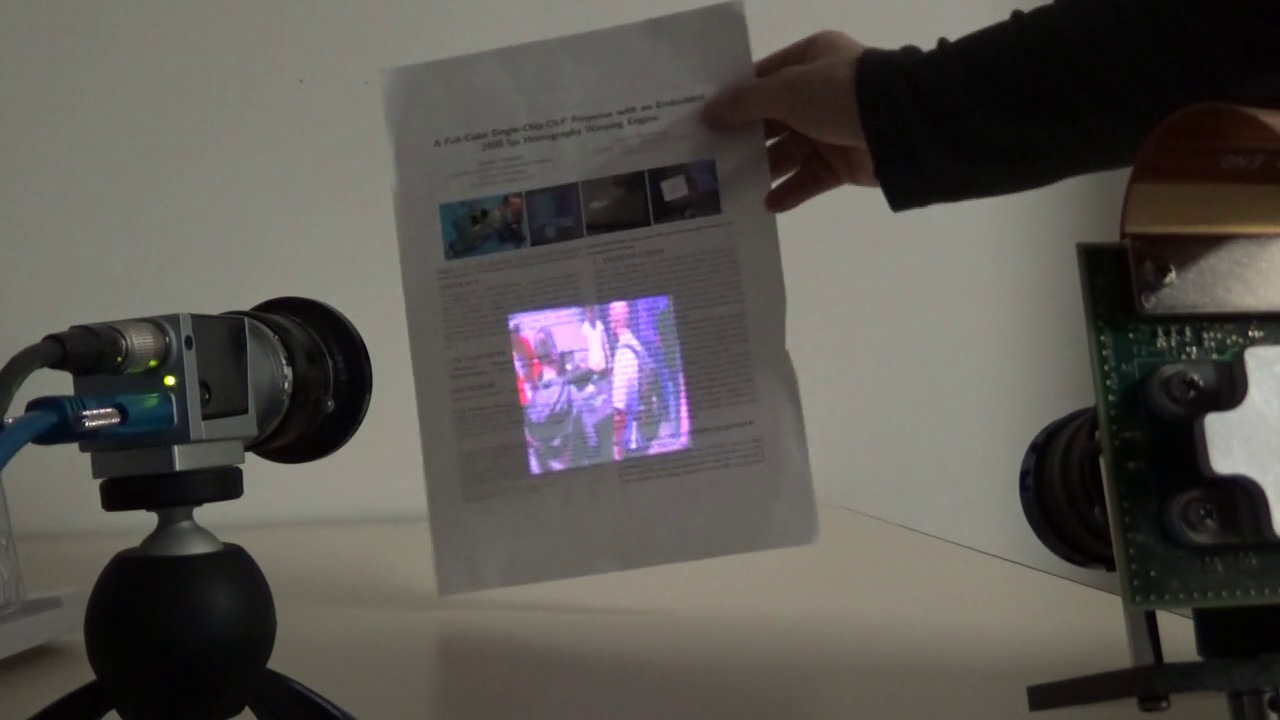}}~%
\subfigure[Animation attachment to a book]{
\includegraphics[width=4.4cm]{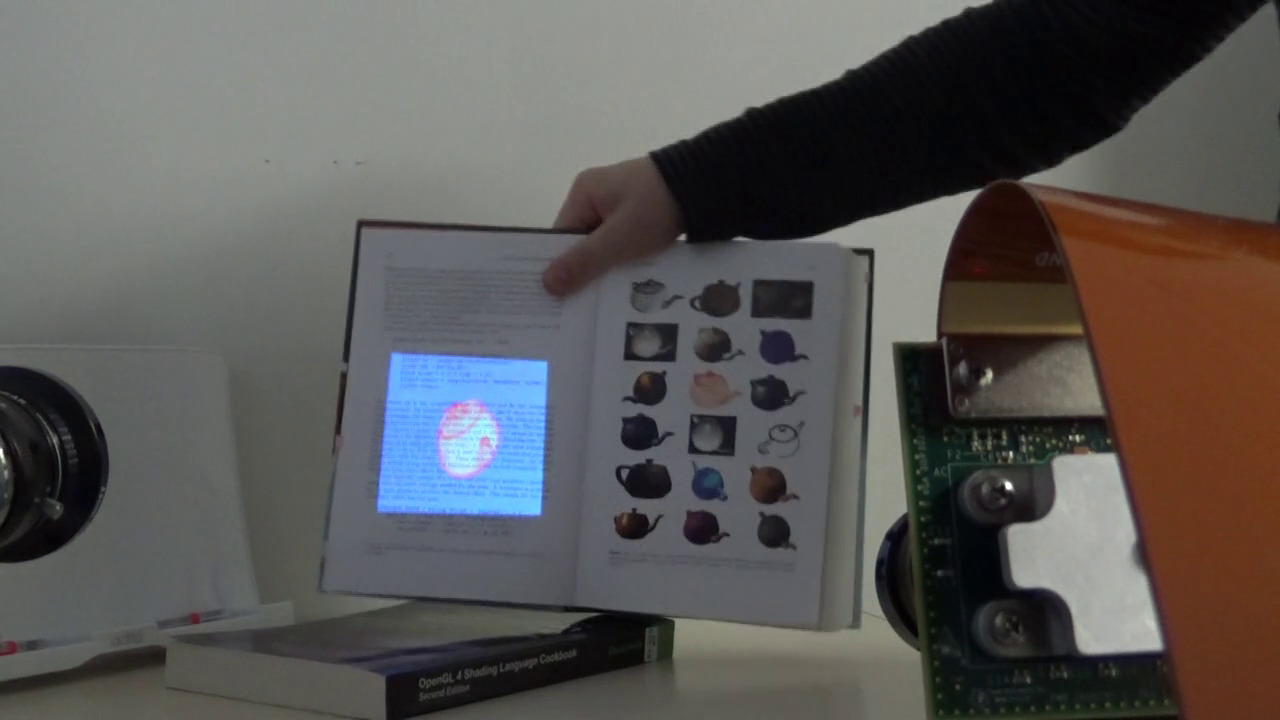}}\\
\subfigure[Annotating a map]{
\includegraphics[width=4.4cm]{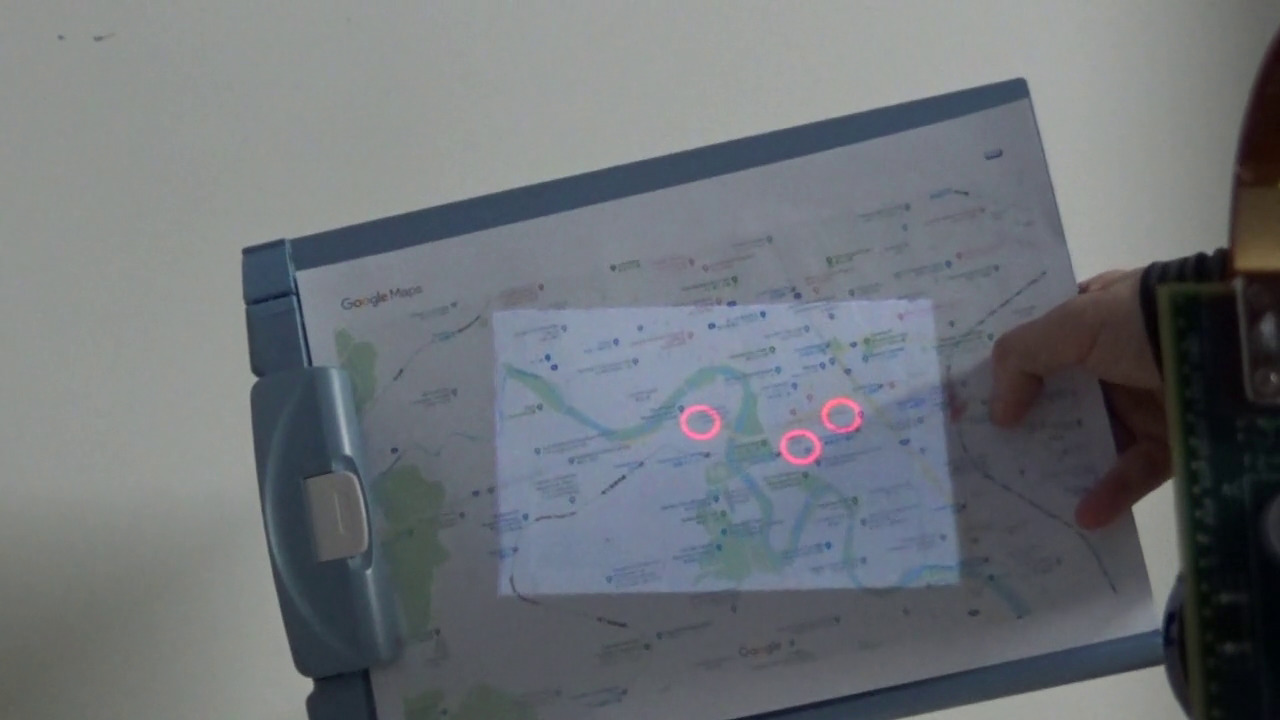}}~%
\subfigure[Visual effect on a guitar]{
\includegraphics[width=4.4cm]{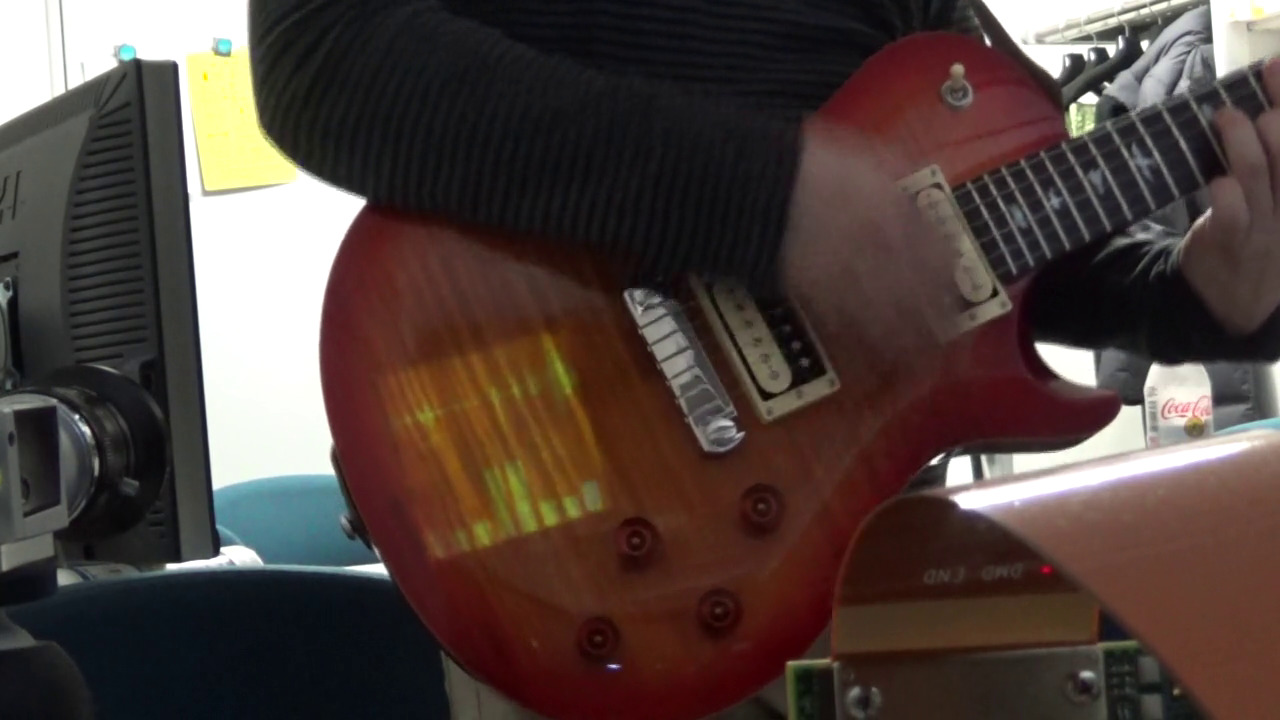}}
\caption{Snapshots of application-suggestive examples.}
\label{fig:snapshots}
\end{figure}

We also present several application-suggestive examples of video
content mapping in \autoref{fig:snapshots}. \autoref{fig:snapshots}
(a) and (b) show the results of sticking a related video clip and a 3D
animation clip onto printed research paper and a book page,
respectively. \autoref{fig:snapshots} (c) shows an example of
annotating a map using anchors. Through user interaction, a new anchor mark
can be placed on a surface point, and the anchors remain stuck to the
point even if the surface moves.

Finally, \autoref{fig:snapshots} (d) shows an example of projection
mapping on a guitar body. The waveform and spectrum of live sound
retrieved by a PC microphone were visualized as mapped content. This
demonstration 
exemplifies a limitation of the proposed approach used to carry out markerless surface tracking.
Because of the challenging
nature of the guitar body surface (maple veneer coated with
polyurethane) as a visual tracking target, the tracking was not
satisfactorily accurate and was prone to tracking loss.

\section{Limitations and Discussion}
\label{sec:limitations}

\subsection{Impact on Content Visibility}

The proposed approach compromises the contrast of the visible video
content through the insertion of fiducial patterns. The decrease in the
highest brightness owing to this compromise can be overcome using stronger light
sources, although the increase in the black level may occasionally limit
the content design. This is significantly visible in
\autoref{fig:snapshots}~(c), where the solid-color rectangle is not
a part of the content but the result of the increase in black level. This
effect can be suppressed by lowering the light source intensity 
only when the fiducial binary frames are illuminated; however, this will
necessitate a higher-sensitivity camera to maintain stable fiducial
tracking.

Another issue stemming from the inserted fiducials is the possibility of
artifact perception. Although na\"{i}ve viewers can barely notice the
existence of fiducials in video content through our implementation, a
viewer having knowledge regarding the underlying mechanism can occasionally
notice the chessboard boundary by looking closely, particularly when
the surface and content are not texture-rich and are close to a solid color.
A possible way to work around these contrast and artifact problems is
to introduce an infrared projector light source instead of a white light, although the
emissions of infrared light may interfere with other peripheral systems
using infrared such as motion capture and communication apparatuses.

\subsection{Dependency on Surface Textures}

Although the dependency of tracking on the video content has been
completely removed, the proposed approach still suffers inherently
from a dependency on the surface textures. This approach is of course not applicable
to a solid-color surface with no texture. Note, however, that it is
possible to combine the methods assuming a solid-color surface
and a known shape
\cite{johnson2007, resch2016, kagami2015} with the imperceptible
fiducial pattern approach, which is rather easy to achieve.

Even if the surface textures are available, because
only a subset of pixels is used for surface tracking, we should note
that there can be
unfortunate situations in which the pixels chosen for tracking do
not offer sufficiently rich textures.

\subsection{Assuming a Planar Target}

This paper has focused on tracking a projection onto a planar surface.
Generalization to non-planar surfaces is not impossible because
direct alignment methods in general are applicable to non-planar
objects. For example, an extension to the ESM algorithm has been
proposed~\cite{silveira2010}. However, we should note that the
proposed method uses a subset of pixels for optimization. The surface
points at the unchosen pixels are not measured, and we therefore must rely on interpolation. For unknown or deformable shapes, we will
require extra assumptions such as the spatial smoothness or known dynamics.

The fiducial patterns may also need to be redesigned,
particularly when the surface shape is complicated, because the
trackability of a pattern with a direct alignment method is affected
by the distortion of the pattern observed from the camera view, and
the simple pattern design used in this paper may not be sufficient.

\subsection{Necessity for Special Hardware}

The proposed approach is built upon the availability of a high-speed
projector-camera pair. This is partly in order to ensure a quick
adaptation to fast motion and to enable the tracking algorithm to
operate by keeping the image displacement between consecutive frames small.
The necessity for special hardware may sound limiting, but we believe
that low-latency camera-projector feedback is essentially required for
dynamic projection mapping applications.

It should be noted that the proposed algorithm can be used with
different types of high-speed projection systems if they have
a real-time rendering capability for high frame rate images as well as
synchronization with the camera, although we believe the architecture
adopted in our implementation is well suited for consumer-level
applications.

\section{Conclusion}

This paper described an approach to achieve a fast projection mapping of
video content onto a markerless planar surface using an uncalibrated
projector-camera pair. A closed-loop alignment has been achieved by
inserting fiducial patterns into the binary frame sequence of a DMD
projector, which are designed to enable surface tracking and fiducial
tracking simultaneously from a single camera image. It was found that
400-fps visual feedback control with the compensation of a one camera frame
delay, without modeling the target dynamics, works well in enabling a
quick adaptation of content onto fast moving surfaces. 
In a future
study, we aim to tackle some of the limitations described in
\autoref{sec:limitations}, including an extension to non-planar surfaces
and an improvement of the content visibility by redesigning the fiducial
patterns when considering human visual characteristics.

\acknowledgments{Part of this work was supported by JST ACCEL
JPMJAC1601 and JSPS Grant-in-Aid 19H04146, 16H02853, and 16H06536.}

\bibliographystyle{abbrv-doi}

\bibliography{ismar2019_kagami_arxiv}
\end{document}